\documentstyle[floats,aps,epsf,amsfonts]{revtex}
\def\lesssim{\mathrel{\hbox{\rlap{\hbox{\lower4pt\hbox{$\sim$}}}\hbox{$<$}}}}
\def\gtrsim{\mathrel{\hbox{\rlap{\hbox{\lower4pt\hbox{$\sim$}}}\hbox{$>$}}}}
\newcommand {\be}       {\begin{equation}}
\newcommand {\ee}       {\end{equation}}
\newcommand {\ban} {\begin{eqnarray}}
\newcommand {\ean} {\end{eqnarray}}
\tighten
\begin{document}
\draft

\title{Confusion Noise from LISA Capture Sources}


\author{Leor Barack$^{1,2}$ and Curt Cutler$^3$}
\address
{$^1$Department of Physics and Astronomy and
Center for Gravitational Wave Astronomy,
University of Texas at Brownsville,
Brownsville, Texas 78520\\
$^2$Department of Mathematics, University of Southampton, Southampton,
SO17 1BJ, United Kingdom\\
$^3$Max-Planck-Institut f\"{u}r Gravitationsphysik,
Albert-Einstein-Institut, Am M\"{u}hlenberg 1,
D-14476 Golm bei Potsdam, Germany}

\date{\today}
\maketitle

\begin{abstract}

Captures of compact objects (COs) by massive 
black holes (MBHs) in galactic nuclei will be an important source for LISA,
the proposed space-based gravitational wave (GW) detector.
However, a large fraction of captures will not be individually
resolvable---either because they are too distant, have unfavorable
orientation,
or have too many years to go before final plunge---and so will constitute
a source of ``confusion noise,'' obscuring other types of sources.
In this paper we estimate the shape and overall magnitude of
the GW background energy spectrum generated by CO captures. This
energy spectrum immediately translates to
a spectral density ${\cal S}^{\rm capt}_h(f)$ for the amplitude of
capture-generated GWs registered by LISA.
The overall magnitude of ${\cal S}^{\rm capt}_h(f)$
is linear in the CO capture rates, which are rather uncertain; therefore
we present results for a plausible range of rates.
${\cal S}^{\rm capt}_h(f)$ includes the contributions from both
resolvable and unresolvable captures, and thus
represents an upper limit on the confusion noise level.
We then estimate what fraction of ${\cal S}^{\rm capt}_h(f)$ is due to
unresolvable sources and hence constitutes confusion noise.
We find that almost all of the contribution to ${\cal S}^{\rm capt}_h(f)$ coming
from white-dwarf and neutron-star captures, and
at least $\sim 30\%$ of the contribution from black hole captures, is from
sources that cannot be individually resolved.
Nevertheless, we show that the impact of capture confusion noise
on the total LISA noise curve ranges from insignificant to modest,
depending on the rates.
Capture rates at the high end of estimated ranges
would raise LISA's overall (effective) noise level
$[f S^{\rm eff}_h(f)]^{1/2}$ by at most a factor $\sim 2$
in the frequency range $1-10$ mHz, where LISA is most sensitive.
While this slightly elevated noise level would
somewhat decrease LISA's sensitivity to {\it other} classes of sources,
we argue that, overall, this would be a pleasant
problem for LISA to have: It would also imply that
detection rates for CO captures were at nearly their maximum possible levels
(given LISA's baseline design and the level of confusion noise from galactic
white-dwarf binaries).

This paper also contains, as intermediate steps, several results
that should be useful in further studies of LISA capture sources, including
(i) a calculation of the total GW energy output from generic inspirals
of COs into Kerr MBHs,
(ii) an approximate GW energy spectrum for a typical capture, and
(iii) an estimate showing that in the population of detected capture sources,
roughly half the white dwarfs and a third of the neutron stars will be
detected when they still have $\gtrsim 10$ years to go before final plunge.

\end{abstract}


\section{Introduction}
\label{Sec:Int}

Captures of stellar-mass compact objects (COs) by massive ($\sim 10^6  M_\odot$)
black holes (MBHs) in galactic
nuclei will be an important source for LISA, the proposed space-based
gravitational-wave (GW) detector.
Recent estimates predict that LISA will detect hundreds to thousands of
such captures over its projected 3--5 year mission lifetime, with
captures of $\sim 10 M_{\odot}$ black holes (BHs) dominating the detection
rate~\cite{emri}. In this paper, we point out that most captures of white dwarfs
(WDs) and
neutron stars (NSs), and a substantial fraction of BH captures, will not be
individually resolvable, and hence will constitute a source of ``confusion
noise,'' obscuring other types of sources.
We estimate the shape and overall magnitude of the GW energy
spectrum due to captures, from which we derive
the spectral density ${\cal S}^{\rm capt}_h(f)$ for the amplitude of
capture-generated GWs registered by LISA.
We then estimate what fraction of ${\cal S}^{\rm capt}_h(f)$
comes from unresolvable sources, and so represents confusion noise.

The plan of this paper is as follows. In Sec.\ \ref{SecII} we briefly review
what is known about the astronomical capture rates.
In Sec.\ \ref{SecIII} we calculate the
total GW energy released by a single capture and estimate its spectrum.
From this, we abstract a model
``average'' capture spectrum (where the average is over
the orbit's final eccentricity and the inclination angle $\iota$
between the MBH's spin and the CO's orbital angular momentum), for
a given MBH mass.
We then further average this model spectrum over MBH mass
(weighted by the capture rate) to estimate the shape of
the GW energy spectrum from all capture sources.
(In practice, we shall only calculate the spectrum in the frequency interval
most relevant to us:  $\sim 1- 10\,$mHz, near the
bottom of the LISA noise curve.)

In Sec.\ \ref{SecIV}  we combine this spectral shape with rate estimates
to produce ${\cal S}^{\rm capt}_h(f)$, the spectrum of GWs
registered by LISA due to all capture sources in the universe.
We calculate ${\cal S}^{\rm capt}_h(f)$ in two passes: First we make
a crude estimate, using
a flat-spacetime cosmology and no source evolution, except for a
sharp cut-off at early times. Then we incorporate cosmological effects
and some simple guesses as to the source evolution, leading to results
that differ little from our initial flat-spacetime estimate.
The overall magnitude of ${\cal S}^{\rm capt}_h(f)$ is linear in the
capture rates; these are rather uncertain, and so we consider a range
of rates taken from the literature.

In Sec.\ \ref{SecV} we address the important issue of {\em source subtraction}:
We ask how much of the capture background calculated in
Sec.\ \ref{SecIV} can be ``fitted out'' (by subtracting
individual contributions from the brightest sources).
The GWs from sources that {\em cannot} be individually resolved, we regard
as confusion noise.
After discussing some general scaling rules between source rates,
confusion noise, and detection rates, we argue that most of
the WD and NS contributions to the capture background, and
at least $\sim 30 \%$ of the BH contribution, is unresolvable.
We shall see that much of the confusion noise comes from sources
that have $\sim 10-200$ years to go before final plunge. Because the
orbits of such sources are still highly eccentric, they radiate a substantial
amount of energy into the
LISA band near perihelion passage; yet, their overall signal-to-noise (SNR)
output is too low for LISA to detect them in a 3-yr integration time.
We refer to these capture sources with $\gtrsim 10$ years yet to live
as ``holding-pattern objects'' (HPOs), to distinguish them from COs now
approaching their ``terminal descent.''
Of course, the nearest HPOs will be detectable,
and HPOs should account for roughly half the
WD-capture detection rate---which was not previously realized.

Finally, in Sec.\ \ref{SecVI} we estimate the
effect of capture confusion noise on the total effective
LISA noise level $[f S^{\rm eff}_h(f)]^{1/2}$.
(This is somewhat complicated by the fact that
capture confusion noise does not simply add in quadrature to the
other noise sources. Roughly, this is because,
in the crucial $2-5\,$mHz band near the
floor of the total noise curve, LISA's effective noise level
is dominated by ``imperfectly subtracted'' galactic binaries, which
effectively reduce the bandwidth available for carrying information
about other types of sources. To a first approximation, the capture
confusion noise gets magnified by a factor that accounts for the lost
bandwidth.)
We conclude that the effect of capture confusion noise on the total LISA
noise curve is rather modest: Even for the highest capture rates
we consider, the total LISA noise
level $[f S^{\rm eff}_h(f)]^{1/2}$ is raised by a factor
$\lesssim 2 $ in the frequency range $1-5\,$mHz.
While this slightly elevated noise level would
somewhat decrease LISA's sensitivity
to {\it other} classes of sources (e.g., the
merger of a $10^4 M_{\odot}$ BH with a  $10^5 M_{\odot}$ BH at $z=1$),
we argue that, overall, this would be a pleasant
problem for LISA to have, since it would be accompanied by
CO-capture detection rates at nearly their maximum possible levels
(given LISA's baseline design and the level of confusion noise from
galactic binaries).

For simplicity, we shall lump COs into three classes: $10 M_{\odot}$ BHs,
$1.4 M_{\odot}$ NSs, and $0.6 M_{\odot}$ WDs. (While these fiducial mass values
are unobjectionable for NSs and accurate to within a factor $\sim 2$ for WDs,
it may be a poor approximation to assume all BHs weigh $10 M_{\odot}$.
However, since the distribution function for BH masses is still poorly constrained
by observation~\cite{Orosz_2002}, we have opted here for simplicity.)
We shall also approximate the average
GW spectra from these three classes as having identical shapes.
In reality,
since the distribution
of initial conditions immediately after capture (especially, the
distribution of initial pericenter) are presumably somewhat
different for the three classes, the average spectra
must also be somewhat different.
However, since good models of these distribution
functions are currently not available (and since our ``average spectrum''
is anyhow just a crude approximation), we again opt for simplicity
and neglect any such differences between our three classes of COs.

We use units in which $G=c=1$.  Therefore, everything can be measured in
the fundamental unit of seconds. However, for the sake of familiarity,
we also sometimes express quantities in terms of yr, Gpc, or $M_\odot$,
which are related to our fundamental unit by 1 yr $= 3.1556 \times 10^7$s,
1 Gpc $= 1.029 \times 10^{17}$s, and $1 M_\odot = 4.926 \times 10^{-6}$s.

\section{Summary of capture rates and source parameter distributions}
\label{SecII}

Here we summarize the astronomical inputs necessary for our
analysis: current estimates of capture rates for the three
CO types and estimated distributions of source parameters.
We refer to Ref.\ \cite{emri} for a more extended discussion;
here we mainly just quote the relevant results.

Captures occur when two objects in the dense stellar cusp surrounding a galactic
MBH undergo a close encounter, sending one of them into an orbit
tight enough that orbital decay through emission of gravitational radiation
dominates the subsequent evolution. For a typical capture, the
initial orbital eccentricity is extremely large
(typically $1-e\sim 10^{-6}{-}10^{-3}$) and the initial pericenter
distance very small ($r_{\rm p}\sim 8-100M$, where $M$ is the MBH mass)
~\cite{FreitagApJ}.
The subsequent orbital evolution may be divided into three stages
(this division is more qualitative than strict):
In the first and longest stage, the orbit is extremely eccentric, and GWs
are emitted in short ``pulses'' during pericenter passages. These GW pulses slowly
remove energy and angular momentum from the system, and the orbit gradually
shrinks and circularizes. After $\sim 10^3-10^8$ years (depending on the
two masses and the initial eccentricity---see Eq.\ (1) of Ref.\
\cite{FreitagApJ}),
the evolution enters
its second stage, when the orbit is sufficiently circular that the emission can
be viewed as continuous.
Finally, the adiabatic inspiral transitions to a direct plunge, as the object
reaches the last stable orbit (LSO). In this final, very brief stage, the
object quickly
plunges through the MBH's horizon, and the GW signal cuts off.
While individually-resolvable captures will mostly be detectable during
the last $\sim 1-100$ yrs of the second stage (depending on the CO and MBH
masses), radiation emitted during the first stage (mostly in short bursts near
periastron passage) will contribute significantly to the confusion background.

One of our goals is to estimate the ambient GW energy spectrum
arising from all captures in the history of the universe.
To this end, we shall
need the following pieces of information:
(i) the space density and mass distribution of MBHs;
(ii) the distribution of MBH spins;
(iii) the capture rate per MBH as a function of the CO mass; and
(iv) the distribution of plunge eccentricities.
In the following we summarize current estimates for these quantities.

\paragraph*{Number density of MBHs:}
We shall see that, for captures by a MBH of mass $M$, the GW energy spectrum is peaked
near frequencies $f \sim 5\, M_6^{-1}\, (1+z)^{-1}\, {\rm mHz}$,
where $M_6\equiv M/10^6 M_{\odot}$ and $z$ is the cosmological
redshift at emission.
The space density of MBHs, as a function of MBH mass $M$, has been
estimated by Aller and Richstone~\cite{AR}, using the measured
correlations between $M$ and the  bulge velocity dispersion $\sigma$
($M \propto \sigma^5 $) in relatively nearby galaxies, as well as the
measured correlation between $\sigma$ and galactic luminosity.
Here we are principally
interested in GWs that might effectively raise the floor of the LISA
noise curve, in the range $\sim 1-10$ mHz.
Therefore we will restrict attention to MBH masses
in the range $0.1\leq M_6\leq 10$.
For $M < 10^7\, M_{\odot}$, the space
density of MBHs, per logarithmic mass interval, turns out to be nearly
$M$-independent and is given by~\cite{AR}
\begin{equation} \label{numberdensity}
dN/d\log M= 2\times 10^{6}\,\gamma\, h_{70}^2\,{\rm Gpc}^{-3},
\end{equation}
where $h_{70}=H_0/(70$ km s$^{-1}$ Mpc$^{-1})$ and $\gamma$ is a
number of order one.
The Aller and Richstone  distribution corresponds to $\gamma\sim 2$;
however, if Sc-Sd galaxies are removed from the sample [as at least some
of them (e.g., M33 and NGC 4395) have MBH masses much lower than
would be predicted from the host galaxy's luminosity], this would
produce a more conservative estimate of $\gamma\sim 1$ \cite{emri}.
We shall retain $\gamma$ as an unknown factor of order unity.

\paragraph*{MBHs' spin:}
Theoretically, the spins of MBHs may take any value between zero
and $M^2$. The actual values of astrophysical MBH spins are difficult
to measure directly, and remain the subject of much debate. However,
recent arguments suggest high spin values: $J \sim 0.7-0.95 M^2$
~\cite{spin}.

\paragraph*{Capture rates:}
In order to survive tidal disruption before or during the final inspiral, the
captured object must be compact: either a WD,
a NS or a BH. Let us denote by ${\cal R}^{\rm A}(M;0)$
the present-day capture rate (number of captures per unit proper time
per galaxy) of CO species `A' by a MBH of mass $M$, where
`A' stands for either WD, NS,
or (stellar-mass) BH. Freitag's simulations of the
Milky Way~\cite{Freitag} predicted present day capture rates of
${\cal R}^{\rm WD}=5\times 10^{-6}$ yr$^{-1}$ and
${\cal R}^{\rm NS}=
{\cal R}^{\rm BH}=10^{-6}$ yr$^{-1}$ in our galaxy (where he assumed
all captured WDs, NSs, and BHs have masses $m=0.6 M_{\odot}$, $1.4 M_{\odot}$,
and $9 M_{\odot}$, respectively).
An extrapolation of these results to other MBH masses yields \cite{emri}
\begin{eqnarray} \label{rate}
{\cal R}^A(M;0)= \kappa\!^{A} M_6^{3/8} {\rm yr}^{-1},
\end{eqnarray}
where the species-dependent coefficients $\kappa^{A}$ are estimated
(from the Freitag~\cite{Freitag} rates) by
$\kappa^{\rm WD}=4\times 10^{-6}$ and
$\kappa^{\rm NS}=\kappa^{\rm BH}=6\times 10^{-7}$.
Other capture-rate estimates, including more recent simulations by
Freitag~\cite{FreitagApJ},
generally predict lower rates---for a survey of rate estimates, see \cite{Sigurdsson_03}.
(In particular, we note that Freitag~\cite{Freitag} did not take account of
NS natal kicks, which might deplete the population of NSs in the
cusp by a factor $\sim 10$~\cite{Sigurdsson_03}.)
Consequently, one must allow for quite large uncertainties in the
above rates: More conservative estimates would be $100$ times lower
for $\kappa^{\rm WD}$ and
$10$ times lower for $\kappa^{\rm NS}$ and $\kappa^{\rm BH}$~\cite{emri}.
Hence, in our analysis we will explore the effect of $\kappa$'s in the ranges
\begin{eqnarray} \label{kappa}
\lefteqn{4\times 10^{-8}\leq \kappa^{\rm WD}\leq 4\times 10^{-6},}\nonumber\\
&& 6\times 10^{-8}\leq \kappa^{\rm NS},\kappa^{\rm BH}\leq 6\times 10^{-7}.
\end{eqnarray}

\paragraph*{Plunge eccentricities:}
Strong-field radiation reaction rapidly circularizes the CO's trajectory
(see, e.g.\ Figs.\ 7 and 8 of \cite{BC}). However, captured COs are
initially scattered into orbits with such high eccentricity that
most retain moderate eccentricity all the way to the final plunge.
Based on Freitag's simulations \cite{FreitagApJ} we estimated \cite{BC}
that roughly half the captured $10 M_{\odot}$ BHs plunge with eccentricity
larger than 0.2. Freitag's results (see Fig.\ 1 of \cite{FreitagApJ}) suggest
that WDs initially scatter, and finally plunge, with somewhat smaller
eccentricities, but we found it hard to quantify this effect.

\section{GW energy spectrum from captures}
\label{SecIII}
\subsection{Total GW energy emitted in a single capture} \label{subsecIIIA}

It is a simple problem in classical general relativity to calculate the
total energy radiated by a particle spiralling into a Kerr black hole
on an arbitrary (inclined, eccentric) orbit, but to our surprise we could not locate
the result in the literature \cite{Schmidt}. We supply it  here.

Let $\alpha$ be the total energy radiated to infinity in GWs over
an entire given inspiral, expressed as a fraction of the CO's mass, $m$.
We want to determine $\alpha$ as a function of the orbit's eccentricity and
inclination (with respect to the MBH's spin axis) just prior to plunge.
Here we neglect
the GW energy emitted by the CO during the final
(very brief) plunge, since it is smaller than the energy emitted
during the adiabatic inspiral by a factor of order $m/M$.
Also, we neglect the fraction of energy that goes down the black hole:
Hughes~\cite{Hughes} has estimated that fraction as less
than one percent of the total mass in all cases.
Under these assumptions, $\alpha$ is
approximated as
\begin{equation}
\alpha\simeq 1-E/m,
\end{equation}
where $E$ is the energy of the CO at the LSO. 
(Recall that $E$ includes the CO's rest mass, and takes the value of $m$
for a static object at infinity.)

Geodesic orbits around a Kerr BH are specified (modulo three initial
phase angles)
by three ``constants of motion'': the energy $E$, the
``$z$'' component of the angular momentum $L_z$, and the ``Carter
constant'' Q. The ``radial'' equation of motion along each specific
geodesic is given by \cite{MTW}
\begin{eqnarray}
\rho^4 m^2\left(\frac{dr}{d\tau}\right)^2&=&
\left[E(r^2+a^2)-aL_z\right]^2
-\Delta\left[m^2 r^2+(L_z-aE)^2+Q\right] \nonumber\\
&\equiv& R(r;E,L_z,Q),
\end{eqnarray}
where $a$ is the MBH's spin parameter
(i.e., its angular momentum $J$ per $M$),
$\tau$ is the proper time along the orbit,
$\rho^2=r^2+a^2\cos^2\theta$, and $\Delta=r^2-2Mr+a^2$.
($\theta$ and $r$ are the standard Boyer-Lindquist coordinates.)
Due to radiation reaction, $E$, $L_z$, and $Q$ are not strictly
constant, but slowly evolving in time. The actual inspiral orbit
can then be thought of as osculating through a continuous sequence of geodesics.
For given $M$ and $a$, the LSO is encountered
in the first instance that both of the following conditions are met:
\begin{equation} \label{LSO}
R(r) = \frac{dR(r)}{dr} = 0
\end{equation}
(where the derivative is taken with fixed $E$, $L_z$, and $Q$).
For a given MBH (given $M$, $a$), Eqs.\ (\ref{LSO}) yield a 2-parameter
family of solutions. For instance, one can specify $Q_{\rm LSO}$ and $E_{\rm LSO}$,
and then solve the algebraic Eqs.\ (\ref{LSO}) simultaneously for
$(L_z)_{\rm LSO}$ and $r_p$ (pericenter radius at the LSO).\footnote
{Equations (\ref{LSO})
generally admit 10 solution pairs, most of which are unphysical.
Naturally, one has to select those solutions with $r_p$ real and
greater than the event horizon's radius, $r_h=M+\sqrt{M^2-a^2}$.}
That is, in fact, how we proceed.
Next we obtain the apocenter radius $r_a$ as another root of
$R[r;E,(L_z)_{\rm LSO},Q]=0$, and then we use these values to determine the
eccentricity and inclination angle $\iota$ at the LSO:
\begin{equation}
e_{\rm LSO}=\frac{r_a-r_p}{r_a+r_p},  \quad\quad
\cos\iota_{\rm LSO}=\frac{(L_z)_{\rm LSO}}{\sqrt{(L_z)_{\rm LSO}^2+Q}}\,.
\end{equation}
(We adopt here the definition of \cite{Hughes} for the inclination
angle.) We thus obtain $E_{\rm LSO}$ for this particular eccentricity and inclination.
By varying the prescribed $E_{\rm LSO}$ and $Q_{\rm LSO}$, we
can cover the entire physical
($e,\cos\iota$) plane, and hence obtain $E_{\rm LSO}(e_{\rm LSO},\cos\iota_{\rm LSO})$.

\begin{figure}[htb]
\centerline{\epsfysize 8cm \epsfbox{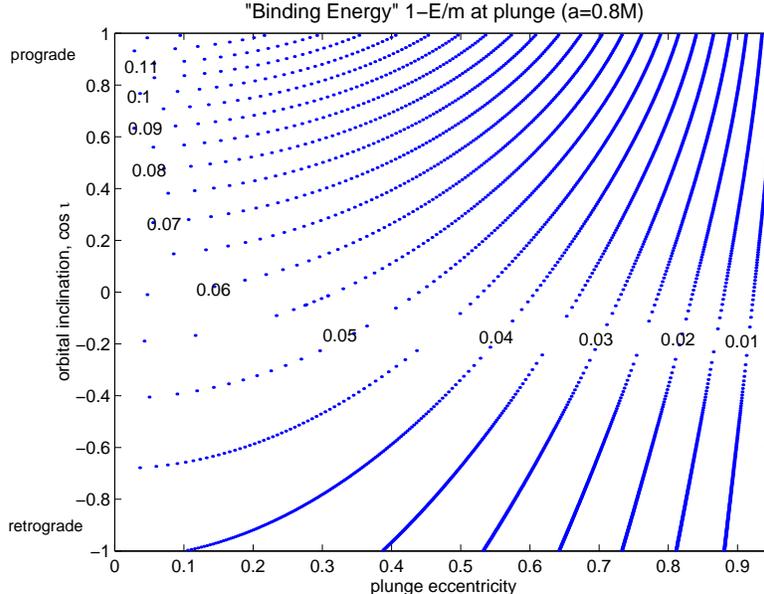}}
\caption{\protect\footnotesize
``Specific binding energy'', $1-E/m$, at plunge, as a function of
the plunge eccentricity and inclination angle, for a Kerr BH with spin parameter
$a=0.8M$. The quantity $1-E/m$ approximates $\alpha$---the amount
of energy (per CO's mass $m$) emitted to infinity in GWs during the entire inspiral.
For the case $a=0.8M$ shown here, and for astrophysically relevant inspirals
with $e_{\rm LSO}\protect\lesssim 0.35$, the CO emits between $\sim 4\%$ and $\sim 12\%$
of its mass in GWs (the former value for retrograde, the latter for prograde
equatorial orbits).
}
\label{fig-alpha}
\end{figure}

\begin{figure}[htb]
\centerline{\epsfysize 7cm \epsfbox{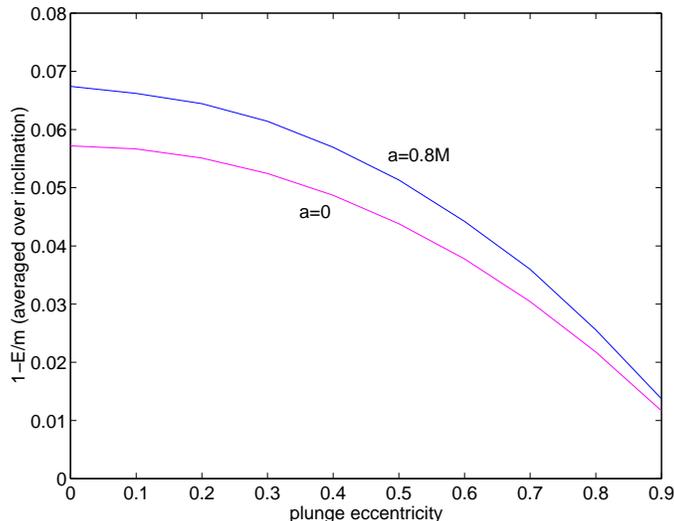}}
\caption{\protect\footnotesize
Specific binding energy at plunge, averaged over (uniformly distributed)
inclination angles, as a function of the plunge eccentricity.
The plot compares the case $a=0.8$ of Fig.\ \ref{fig-alpha},
with the case of a non-spinning MBH ($a/M=0$). In the latter case,
the plunge energy is given by Eq.\ (\ref{Eplunge-Sch}).
An ``inclination averaged'' inspiral in the astrophysically relevant
range of plunge eccentricities ($e_{\rm LSO}\protect\lesssim 0.35$),
emits around 5--7\% of the CO's mass in GWs, with only little dependence
on the MBH's spin.}
\label{fig-alphaAv}
\end{figure}

We implemented the above algorithm using a simple Mathematica script.
Figure~\ref{fig-alpha} shows the results for the case where $a/M=0.8$.
For astrophysically relevant inspirals, with $e_{\rm LSO}\protect\lesssim 0.35~\cite{BC}$,
the CO emits between $\sim 4\%$ and $\sim 12\%$ of its mass in GWs (the
former value for retrograde, the latter for prograde equatorial orbits).
Figure\ \ref{fig-alphaAv} shows the value of $\alpha$ averaged over
inclination angles (assuming a uniform distribution in $\cos\iota$),
as a function of plunge eccentricity. In Fig.~\ref{fig-alphaAv} the case
$a/M=0.8$ is compared
to the Schwarzschild case ($a/M=0$), where the specific energy at plunge is
given analytically by
\begin{equation} \label{Eplunge-Sch}
m^{-1}E_{\rm LSO}=4[(6+2e_{\rm LSO})(3-e_{\rm LSO})]^{-1/2} \quad
\text{(for $a=0$)}
\end{equation}
(see, e.g., Eqs. (2.5) and (2.8) of \cite{Cutler-Kennefick-Poisson}).
Fig.~\ref{fig-alphaAv} shows that the
average (over the inclination angle $\iota$) energy emitted
by inspirals into rapidly rotating BHs is only a little higher than
for inspirals into a Schwarzschild BH.

In the analysis below, we
will adopt $\alpha = 0.06$ as a fiducial average value, based on Fig.~2.

\subsection{GW energy spectrum from a single capture}

Ideally, we would next compute
the spectrum $f(dE/df)$ of GW energy emitted over
the entire inspiral, for given values of $m$, $M$, $a$,  $e_{LSO}$, and
${\iota}_{LSO}$. However, that would be a much harder problem than the one
in Sec.\ \ref{subsecIIIA}. The tools to do a fully satisfactory job
do not even exist yet, since there is still no working code to
calculate the effect of GW radiation reaction on an
arbitrary geodesic orbit in Kerr. Nevertheless, a reasonable job
at approximating the orbital evolution could presumably be done today
by inferring the rate
of change of the CO's $E$ and $L_z$ from
the flux of energy and angular momentum at
infinity,
and approximating the orbital inclination angle $\iota$ as fixed during
the inspiral (Hughes \cite{Hughes} discusses the justification of this
approximation)---which determines the rate of change of $Q$.
Knowing the orbital evolution, one could obtain the GW spectrum by
solving the Teukolsky equation for the emitted GWs.

However, astronomical capture rates are sufficiently uncertain that we feel
justified in a much cruder approach to this problem: We calculate the
spectrum using the approximate, analytic formalism we previously developed
in \cite{BC}. In this formalism, the overall orbit is imagined as osculating
through a sequence of Keplerian orbits, with rate of change of energy,
eccentricity, periastron direction, etc.\ determined by solving post-Newtonian
evolution equations.  In this approximation, the emitted
waveform $h_{\mu\nu}(t)$,
at any instant, is determined by applying the quadrupole formula to an
instantaneous Keplerian orbit. The quadrupole-formula waveform for an
eccentric, Keplerian orbit was derived analytically long ago by Peters and
Matthews \cite{PM}, so these approximate waveforms are relatively easy
to calculate. These waveforms correctly capture the fact that
eccentric orbits have power at all multiples of the orbital frequency
(even when considering only quadrupole radiation, as we do).
We ``cut off'' the waveform when the orbit reaches the point of plunge for a
Schwarzschild MBH, at $r_p = (6 + 2 e_{LSO})/(1 + e_{LSO})$---thus, we are essentially
restricting ourselves to the $a=0$ case. (This is partly because our post-Newtonian
evolution
equations cannot be trusted for prograde orbits in near-extremal Kerr. However,
given the small difference between the $a=0$ and $a = 0.8 M$ cases in Fig.\
\ref{fig-alphaAv}, and given that we will subsequently average the spectra over
$M$, the restriction to $a=0$ here should not appreciably affect the final
average spectrum.)

\begin{figure}[htb]
\centerline{\epsfysize 8cm \epsfbox{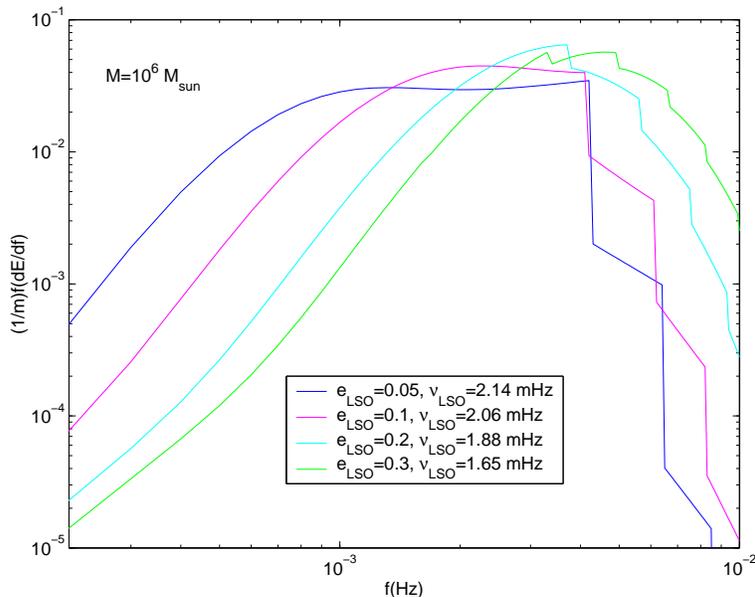}}
\caption{\protect\footnotesize
Energy spectrum of a single inspiral, for $M=10^6 M_{\odot}$, $a=0$, and
for a range of plunge eccentricities $e_{\rm LSO}$. $\nu_{\rm LSO}$ is
the orbital frequency at the LSO. These spectra
are based on approximate, post-Newtonian waveforms as described in the text.
Each of the curves represents the sum of contributions from the first 20 harmonics
of the orbital frequency. Discontinuities appear
whenever an $n$-harmonic reaches frequency $n\times\nu_{\rm LSO}$ and cuts off.
}
\label{fig-spectrum}
\end{figure}

Figure \ref{fig-spectrum} shows the spectrum of a single capture (for
a range of possible plunge eccentricities), as derived from the approximate
waveform model described above. For each value of $e_{\rm LSO}$ we integrated
post-Newtonian evolution equations [Eqs.~(28) and (30) in \cite{BC}]
backward in time to obtain
$e(t)$ and $\nu(t)$, and, consequently, $e(\nu)$. We then obtained the power
radiated into each of the harmonics of the orbital frequency using the
leading-order formula \cite{PM}
\be\label{dotEn}
\dot E_n (\nu)= \frac{32}{5} \mu^2M^{4/3}(2\pi \nu)^{10/3} g_n[e(\nu)],
\ee
where $g_n(e)$ is given by
\ban\label{gne}
g_n(e) &=& \frac{n^4}{32}\bigl\{\bigl[J_{n-2}(ne) - 2e\,J_{n-1}(ne)
+\frac{2}{n}\,J_{n}(ne) + 2e\,J_{n+1}(ne) - J_{n+2}(ne)\bigr]^2
\nonumber \\
&& + (1-e^2)[J_{n-2}(ne) - 2\,J_{n}(ne) + J_{n+2}(ne)]^2
+\frac{4}{3n^2}[J_{n}(ne)]^2 \bigr\}.
\ean
We next re-expressed $\dot E_n(\nu)$ in terms of the GW frequency $f$ by
replacing $\nu\to f_n/n$ for each of the $n$-harmonics. Finally, we summed
up the power from all harmonics (holding $f$ fixed), and plotted the total
power vs.\ $f$. (In practice, we summed over the first $20$ harmonics.
Note that different harmonics ``cut off'' at different GW frequency,
$f_{\rm LSO}=n\,\nu_{\rm LSO}$, giving rise to the ``discontinuities'' apparent
in Fig.~\ref{fig-spectrum}.)
During the inspiral the source evolves significantly in both
frequency and eccentricity, with the GW power distribution shifting gradually
from high harmonics of the orbital frequency to lower harmonics. This leads
to a spectrum with a steep rise followed by a ``plateau'', as manifested
in the Figure. For comparison, a decaying, quasi-circular orbit would yield
a simple power-law spectrum, $f(dE/df)\propto f^{2/3}$.

\subsection{Model capture spectrum, averaged over $e_{\rm LSO}$}

Ideally, the next step in our analyis would be to integrate the individual
capture spectra over the actual distribution of final eccentricities, in order
to obtain an appropriately weighted average. However, since that distribution
is poorly known, and since our individual spectra are anyway
approximated, we instead introduce the following model for the
``average'' spectrum (for a given MBH mass), suggested by
Fig.\ \ref{fig-spectrum}:
\begin{equation} \label{spectrum1}
f(dE/df)=m\times\left\{
\begin{array}{ll}
(f/f_0)^3,                & f<f_p,       \\
{\rm const}=(f_p/f_0)^3,  & f_p<f<4 f_p, \\
0,                        & f>4f_p.
\end{array}
\right.
\end{equation}
Here, $f_0$ is a normalization parameter we will determine soon, and $f_p$
is the plunge frequency at $e_{\rm LSO}=0$:
\begin{equation} \label{fp}
f_p=\left(2\pi 6^{3/2}M\right)^{-1}=2.20\,{\rm mHz}\times M_6^{-1},
\end{equation}
where, recall, $M_6\equiv M/(10^6M_{\odot})$.

\begin{figure}[htb]
\centerline{\epsfysize 6cm \epsfbox{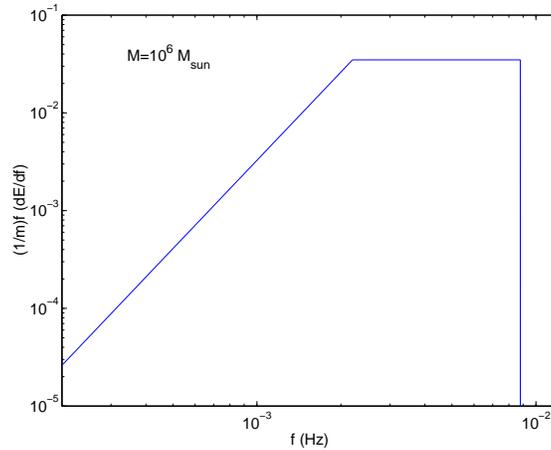}}
\caption{\protect\footnotesize
Our model spectrum [Eq.~(\ref{spectrum2})] for a single
inspiral with $M=10^6 M_{\odot}$, averaged over inclination angle and final
eccentricity, and normalized to $\alpha=0.06$.
The $n^{th}$ GW harmonic cuts off when $f/n$ reaches the plunge
frequency, $f_p$, causing the spectrum to flatten out for
$f_p \protect\lesssim f\protect\lesssim 4f_p$.
We cut off our model spectrum at $f> 4f_p$, where we
no longer find emission from any of the dominant harmonics ($n\sim 2-4$).
}
\label{fig-modelspectrum}
\end{figure}

The parameter $f_0$ is determined from the total energy radiated
in GWs over the entire inspiral. Let this total energy be $\alpha m$, where,
recall, $\alpha$ is roughly our fiducial value of $0.06$.
We have
\begin{equation} \label{intf}
\alpha=m^{-1}\int_0^{\infty}f(dE/df)d\ln f=
\int_0^{f_p} (f/f_0)^3 df/f+\int_{f_p}^{4f_p} (f_p/f_0)^3 df/f=
(f_p/f_0)^3[1/3+\ln(4)],
\end{equation}
and solving for $f_0$ yields
\begin{equation} \label{f0}
f_0=6.74\,{\rm mHz}\times (\alpha/0.06)^{-1/3}M_6^{-1}.
\end{equation}
Thus the energy spectrum can be rewritten as
\begin{equation} \label{spectrum2}
f(dE/df)=(m/M_{\odot})M_{\odot}\times (\alpha/0.06)\times\left\{
\begin{array}{ll}
3.27\times 10^{-3}M_6^3 f_{\rm mHz}^3, & f<f_p,       \\
3.49\times 10^{-2},        & f_p<f<4 f_p, \\
0,                        & f>4f_p\, ,
\end{array}
\right.
\end{equation}
\noindent where $f_{\rm mHz}\equiv f/(1\,{\rm mHz})$.
This spectrum is illustrated in Fig.\ \ref{fig-modelspectrum}.
Our model spectrum is not
particularly accurate, but it is good enough for our purposes: We
will further average it over the MBH mass (in section III.B.), and the
result of this averaging will anyway ``smear out'' the exact details
of the shape shown in Fig.\ \ref{fig-modelspectrum}. For our application,
what matters most is that our model spectrum is peaked at roughly the right
frequency, $f \sim 5\, {\rm Hz}/M_6$, that it is relatively narrow
(most of the energy is contained within a factor $\sim 5-10$ range in
frequency), and that it ``contains'' the right total amount of energy.

\subsection{Model capture spectrum, averaged over MBH mass}

The above model spectrum corresponds to a given MBH mass $M$.
We next average this spectrum over $M$, with the average being weighted
by the capture rate for MBHs of mass $M$.

From Eq.\ (\ref{rate}), the (current) rate of captures by MBHs of mass $M$
is $\propto M^{3/8}$.
The space number density of MBHs with masses $0.1\lesssim M_6<10$ is approximately
constant in this range [see Eq.\ (\ref{numberdensity})].
Therefore, the average GW energy spectrum from a single capture
(per unit mass of the CO) is approximately
\be \label{avespect}
\epsilon(f) = 0.1923\, (\alpha/0.06)\, \int_{0.1}^{10} (dM_6/M_6) M_6^{3/8}
\times \left\{
\begin{array}{ll}
3.27\times 10^{-3}M_6^3 f_{\rm mHz}^3, & f<f_p,       \\
3.49\times 10^{-2},        & f_p<f<4 f_p, \\
0,                        & f>4f_p,
\end{array}\right.
\ee
where $f_{p}\equiv f_p(M_6)$ is given in Eq.\ (\ref{fp}), and
the prefactor 0.1923 is just $[\int_{0.1}^{10} (dM_6/M_6) M_6^{3/8}]^{-1}$.

To evaluate the integral in Eq.\ (\ref{avespect}) we consider
separately 5 different frequency ranges:\\
{\bf For $\mathbf{f<0.22}$ mHz} we have $f<f_p$ for any $M$ in the
integration domain. This low-frequency range is therefore controlled
by the ``tail'' part of the spectrum, and we find
\begin{eqnarray} \label{band1}
\epsilon(f<0.22\,{\rm mHz})&=&
(\alpha/0.06)\, 6.297\times 10^{-4}f_{\rm mHz}^3
\int_{0.1}^{10} dM_6 M_6^{19/8}\nonumber\\
&=&(\alpha/0.06)\, 4.425\times 10^{-1}f_{\rm mHz}^3 \, .
\end{eqnarray}
{\bf For $\mathbf{0.22<f<0.88}$ mHz} we have $f<4f_p$ for any $M$,
and now there are both ``tail'' and ``plateau'' contributions:
\begin{eqnarray} \label{band2}
\epsilon(0.22<f<0.88\,{\rm mHz})&=&
(\alpha/0.06)\left[
6.297\times 10^{-4}f_{\rm mHz}^3\int_{0.1}^{2.2/f_{\rm mHz}}M_6^{19/8}dM_6+
6.710\times 10^{-3}\int_{2.2/f_{\rm mHz}}^{10} M^{-5/8}dM_6
\right]
\nonumber\\
&=&(\alpha/0.06)\,4.243 \times 10^{-2}(1-0.5039\,f_{\rm mHz}^{-3/8}).
\end{eqnarray}
{\bf For $\mathbf{0.88<f<22}$ mHz} we need to cut off the integration
at $M_6=8.8\,{\rm mHz}/f$:
\begin{eqnarray} \label{band3}
\epsilon(0.88<f<22\,{\rm mHz})&=&
(\alpha/0.06)\left[
6.297\times 10^{-4}f_{\rm mHz}^3\int_{0.1}^{2.2/f_{\rm mHz}}M_6^{19/8}dM_6+
6.710\times 10^{-3}\int_{2.2/f_{\rm mHz}}^{8.8/f_{\rm mHz}} M^{-5/8}dM_6
\right]
\nonumber\\
&=&(\alpha/0.06)(1.907\times 10^{-2}f_{\rm mHz}^{-3/8}- 7.881 \times 10^{-8}f_{\rm mHz}^{3}).
\end{eqnarray}
{\bf For $\mathbf{22<f<88}$ mHz} there remains only a plateau contribution:
\begin{eqnarray} \label{band4}
\epsilon(22<f<88\,{\rm mHz})&=& (\alpha/0.06)6.710\times 10^{-3}
\int_{0.1}^{8.8/f_{\rm mHz}} M^{-5/8}dM_6 \nonumber\\
&=& (\alpha/0.06)\,4.046 \times 10^{-2} (f_{\rm mHz}^{-3/8}-0.1866).
\end{eqnarray}
Lastly, we have
\begin{equation} \label{band5}
\epsilon(f>88\,{\rm mHz})=0.
\end{equation}

\begin{figure}[htb]
\centerline{\epsfysize 8cm \epsfbox{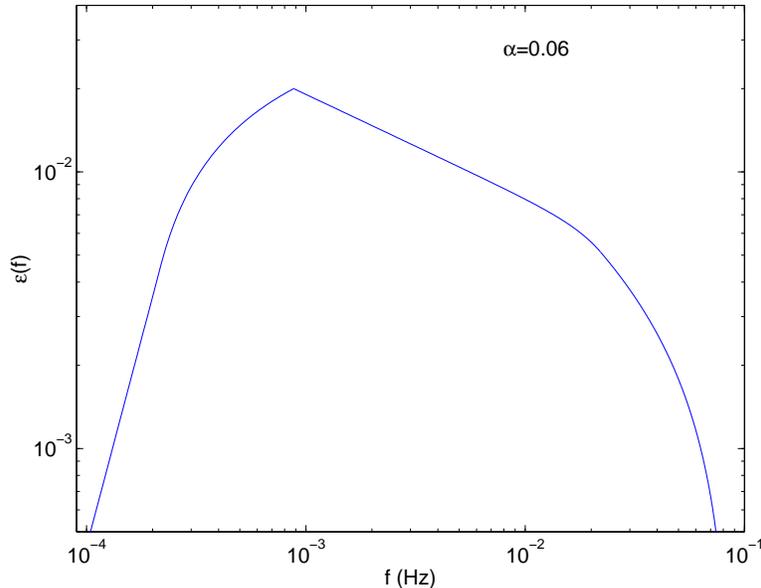}}
\caption{\protect\footnotesize
Model energy spectrum for a single capture, averaged over MBH mass and
normalized by $\alpha=0.06$---see Eq.\ (\ref{epsilon}).
Note the apparent sharp drops in the spectrum
at frequencies below $\sim 10^{-3}$ Hz and above $\sim 2\times 10^{-2}$ Hz are
simply an artifact of our restricting attention to MBH masses in the range
$10^{5}-10^{7} M_{\odot}$.}
\label{fig-backgroundspectrum}
\end{figure}

To summarize, the energy spectrum from captures (per unit CO mass), averaged
over all captures by MBHs with masses in the range $10^5$ to $10^7 M_\odot$, is
\begin{equation} \label{epsilon}
\epsilon(f)= (\alpha/0.06)\times
\left\{
\begin{array}{ll}
4.425\times 10^{-1}f_{\rm mHz}^3,                   & f_{\rm mHz}<0.22,       \\
4.243\times 10^{-2}(1-0.5039\,f_{\rm mHz}^{-3/8}),      & 0.22<f_{\rm mHz}<0.88, \\
1.907\times 10^{-2} f_{\rm mHz}^{-3/8} - 7.881 \times 10^{-8}f_{\rm mHz}^{3},
& 0.88<f_{\rm mHz}>22,\\
4.046\times 10^{-2}(f_{\rm mHz}^{-3/8}-0.1866),       & 22<f_{\rm mHz}<88,\\
0,                               & f_{\rm mHz}>88 \, .
\end{array}
\right.
\end{equation}
This spectrum is depicted in Fig.\ \ref{fig-backgroundspectrum}.

\section{Spectral density of capture background}
\label{SecIV}

\subsection{Total energy spectrum of capture background}

Using the above estimates, we want to determine the ambient GW spectrum
from all captures in the history of the universe.
To simplify the analysis, we will first assume that spacetime is
flat, that the capture rate has been fixed at
${\cal R}^A(0)$ for the last $T= \beta \cdot 10^{10} $ years, and that
previous to that the capture rate had been zero.
In Subsection \ref{Cosmology} below we carry out
a more careful analysis, including cosmological and evolutionary effects,
and show that the above simplified version actually gives reasonably
accurate results, especially if one
makes the judicious choice $\beta = 0.7$.

Using our flat spacetime assumptions,
the GW energy density spectrum from captures
of species A, $d\rho^A_{\rm GW}/d({\rm ln} f)$, is just the average energy spectrum
per capture, $m^A \epsilon(f)$, times the current total capture rate per unit
co-moving volume, ${\cal R}^A(0)$, times $T$. From Eqs.\ (\ref{numberdensity})
and (\ref{rate}), the factor ${\cal R}^A(0)$ is
\begin{eqnarray}
{\cal R}^A(0) &=& \int_{0.1}^{10} {\cal R}^A(M;0) (dN/dM_6) dM_6 \nonumber\\
&=& 1.040\, (\kappa^{A}/10^{-7})\,\gamma\,  h_{70}^2\,{\rm Gpc}^{-3}{\rm yr}^{-1}\, .
\end{eqnarray}
Using $d\rho^A_{\rm GW}/d({\rm ln} f) =  m^A {\cal R}^A(0) T\, \epsilon(f)$, with
the conversion factor
$1 M_\odot/{\rm Gpc}^3 = 4.521 \times 10^{-57}\ {\rm s}^{-2}$ ,
we then find
\begin{eqnarray}\label{drhodf}
\frac{d\rho^A_{\rm GW}}{d({\rm ln} f)}= 3.291 \times 10^{-47}
\left(\frac{m^A}{M_{\odot}}\right)
\left(\frac{\beta}{0.7}\right)
\left(\frac{\kappa^{A}}{10^{-7}}\right)\,\gamma\, h_{70}^2
\times \epsilon(f)\ {\rm sec}^{-2},
\end{eqnarray}
where, recall, $\epsilon(f)$ is given in Eq.\ (\ref{epsilon}) above.

\subsection{LISA noise model} \label{NoiseModel}

Our estimate of $d\rho^A_{\rm GW}/d({\rm ln} f)$ leads directly to
an estimate of the spectral density of the capture background
measured by the LISA detector.
But before proceeding to this, we briefly state our conventions
and review standard
estimates for the magnitudes of {\it other} noise sources.
For more details
we refer the reader to Sec.\ V.A of Ref.\ \cite{BC}.

LISA's GW data stream is basically equivalent to the
output of two synthesized Michelson interferometers, where in
each Michelson the angle between the two arms is $60^{\circ}$, and
where the two Michelsons---hereafter labelled I and II---
are rotated with respect to each other by $45^\circ$.
This representation
is only accurate for low frequencies,
$f \lesssim 0.01$ Hz. However,
most of the SNR from captures will indeed be accumulated at these
lower frequencies, so this approximation is adequate for our purpose.
At low $f$,
our synthesized detectors I and II are basically equivalent to
the data combinations lablelled A and E in the terminology of Time Delay
Interferometry~\cite{Estabrook2000}.
The output $s_I(t)$ of synthetic-Michelson I is the sum of resolvable GW signals $h_I(t)$
and noise $n_I(t)$;
similarly, $s_{II}(t) = n_{II}(t) + h_{II}(t)$.

In keeping with the usual convention in the LISA literature,
we define $S_h(f)$ to be the {\it single-sided, sky-averaged}
noise spectral density
for either synthetic detector, I or II:
\be\label{Sdef}
\langle {\tilde n_I}(f) \, {\tilde n_I}(f^\prime)^* \rangle =
\langle {\tilde n_{II}}(f) \, {\tilde n_{II}}(f^\prime)^* \rangle =
{3\over 40}
\delta(f - f^\prime) S_h(f),
\ee
where a tilde denotes the Fourier transform and ``$\langle\ \rangle$''
stands for the expectation value. The factor ${3\over 40}$ here is the
product of the following three
factors: ${1\over 2}$ from the ``single-sided'' convention,  ${3\over 4} =
{\rm sin}^2 60^{\circ}$ because the angle between the arms is
$60^{\circ}$, and ${1\over 5}$ from an average over source directions
and polarizations.
[Note here we are using a different convention than in our earlier paper~\cite{BC},
where $S_h(f)$ was defined without the sky-averaging factor.\footnote{Note the PRD version
of \cite{BC} also contains errors, as follows: The RHS of Eq.~(48) should be multiplied by
$\frac{3}{4}$; the factor $5$ in Eq.~(55) and in the sentence below it should both be replaced
by $\frac{20}{3}$; and the same for the factor 5 in the three sentences preceding
Eq.~(60).}]

LISA's noise has three main components (besides confusion noise from captures):
instrumental noise, confusion noise from short-period galactic binaries
(mostly WD binaries), and confusion noise from short-period
extragalactic binaries.
For LISA's instrumental noise, $S^{\rm inst}_h(f)$, we use the curve
available from S. Larson's {\it online sensitivity curve generator}
\cite{Larson}, which is based on the noise budgets specified in the LISA
Pre-Phase A Report~\cite{Pre}.
For $f \lesssim 5$ mHz, the instrumental noise curve
is well fitted analytically by \cite{FT}
\be\label{noise_Sum}
S^{\rm inst}_h(f) = 6.12 \times 10^{-51}f^{-4} + 1.06\times 10^{-40}
+ 6.12 \times 10^{-37}f^{2} \ {\rm Hz}^{-1},
\ee
where $f$ is in Hz.

Next we turn to confusion noise from galactic and extragalactic WD binaries
(GWDBs and EGWDBs, respectively).
Any isotropic background of individually unresolvable GW sources represents
(for the purpose of analyzing {\it other} sources) a noise source with
spectral density~\cite{stochUL}~\footnote{Note the RHS of our Eq.~(\ref{ShOm})
is a factor
$5 = (20/3){\rm sin}^2 (\pi/3)$ larger than the RHS in
Eq.~(3.6) in\cite{stochUL}. The factor
$20/3 = 5/{\rm sin}^2 (\pi/3)$ is our usual
conversion factor between LIGO and LISA conventions, and
the ${\rm sin}^2 (\pi/3) = {3\over 4}$ factor arises because a smaller
opening angle reduces LISA's sensitivity to background GWs.
Thus the two factors of ${\rm sin}^2 (\pi/3)$ cancel.
The convention followed
in this paper is such that the ratio of ${\cal S}^{\rm backgd}_h(f)$ to
$d\rho_{\rm GW}/d\ln f$ is independent of the opening angle between the arms.}

\be
\label{ShOm}
{\cal S}^{\rm backgd}_h(f) = \, \frac{4}{\pi} f^{-3}  \frac{d\rho_{\rm GW}}{d({\rm ln} f)} \, .
\ee
Here and throughout this paper, our convention is to use the ``calligraphic''
${\cal S}_h$ to represent the spectral density of the entire background
from some class of sources,
including both resolvable and unresolvable sources.
The fraction
of ${\cal S}_h$ due to the unresolvable sources, which we
refer to as ``confusion noise'', we denote by
$S_h$ with an Italic typeface.
We shall also use the ``upright'' typeface for the
instrumental noise, $S^{\rm inst}_h$.
Estimates of $d\rho_{\rm GW}/d\ln f$ from the galactic and extragalactic
WD backgrounds~\cite{FarmerPhinney,Nelemans_2001c} then yield the following
background spectral densities:
\begin{eqnarray}
{\cal S}^{\rm GWDB}_h(f) &=& 1.4\times10^{-44}\,\left(\frac{f}{1{\rm Hz}}\right)^{-7/3}
{\rm Hz}^{-1}, \\
{\cal S}_h^{\rm EGWDB}(f) &=&
2.8 \times 10^{-46} \left(\frac{f}{1{\rm Hz}}\right)^{-7/3}
{\rm Hz}^{-1}.
\end{eqnarray}
\noindent
(In fact, there is practically no distinction between
${\cal S}_h^{\rm EGWDB}$ and $S_h^{\rm EGWDB}$,
since the {\it entire}
extragalactic contribution is assumed to consist of unresolvable sources.)

The GWDB background is actually larger than LISA's instrumental noise in the
range $\sim 10^{-4}$--$10^{-2}$ Hz. However, at frequencies
$f \agt 2 \times 10^{-3}\,$Hz, galactic sources
are sufficiently sparse, in frequency space, that one expects to be able
to ``fit them out'' of the data.
Still, even resolvable sources introduce an additional kind
of ``noise''
since they can never be subtracted out perfectly.
To the extent
that such ``subtraction errors'' can together mimic other classes of
astrophysically interesting signals, they effectively diminish the
sensitivity of LISA to those other signals.
(This effect is also referred to
simply as ``confusion noise'' within the LISA community, though
conceptually it can be useful to consider the two kinds of
confusion noise separately.)
In the case of GWDBs at frequencies above $\gtrsim 2 $mHz (i.e., at
frequencies where they become individually resolvable), this
second type of confusion noise acts effectively like a multiplicative
factor on all other types of noise. E.g., if one is searching
for coalescing MBH binaries, but the signal from some
fraction $F(f)$ of the
frequency bins near frequency $f$ are rendered effectively unusable for this
purpose because of degeneracies with the fitted GWDBs, then
it is {\it as if} the noise spectral density were increased by
a factor  $(1-F)^{-1}$~\cite{bender_hils_97}.

We denote by $S^{\rm eff}_h$ the total effective noise,
including the multiplicative effect from this second type of confusion noise,
and adopt the following rough
estimate of $S^{\rm eff}_h$, based on Hughes~\cite{Hughes02}:\footnote
{Actually, Hughes~\cite{Hughes02}
writes $S^{\rm eff}_h(f) = {\rm min}
\left\{S_h^{\rm inst}(f) \exp(\kappa T_{\rm mission}^{-1} dN/df), \;\;
S_h^{\rm inst}(f) + {\cal S}_h^{\rm GWDB}(f)\right\}\, + {\cal S}_h^{\rm EGWDB}(f)$, but
it makes no sense to treat ${\cal S}_h^{\rm EGWDB}$ differently from
$S_h^{\rm inst}$ in this context, so we have added them together in Eq.\
(\ref{inst+gal}). However, ${\cal S}_h^{\rm EGWDB}(f)$ is
sufficiently small that this modification has almost no effect in practice.}
\begin{equation}\label{inst+gal}
S^{\rm eff}_h(f)= {\rm min}\left\{\left[S_h^{\rm inst}(f) +
{\cal S}_h^{\rm EGWDB}(f)\right] \exp(\kappa T_{\rm mission}^{-1} dN/df), \;\;
S_h^{\rm inst}(f) + {\cal S}_h^{\rm EGWDB}(f) + {\cal S}_h^{\rm GWDB}(f)
\right\}\, .
\end{equation}
Here $dN/df$ is the number density of GWDBs per unit GW frequency,
$T_{\rm mission}$ is
the LISA mission lifetime (so $\Delta f = 1/T_{\rm mission}$ is the bin size
of the discretely Fourier transformed data), and $\kappa$ is
the average number of frequency bins that are ``lost'' (for the purpose
of analyzing other sources) when each galactic binary is fitted out.
For $dN/df$ we adopt the estimate~\cite{Hughes02}
\begin{equation}\label{eq:dNdf}
{dN\over df} = 2\times10^{-3}\,{\rm Hz}^{-1}\left(1\,{\rm Hz}\over
f\right)^{11/3},
\end{equation}
and take $\kappa T_{\rm mission}^{-1} = 1.5/{\rm yr}$ (corresponding, e.g., to
$T_{\rm mission} \approx 3\,$yr and $\kappa \approx 4.5$).

Even though the subtracted noise model, Eq.~(\ref{inst+gal}),
basically represents an ``educated guess,''
it has the right qualitative features:
The very steep slope of $dN/df$ means that a factor $3$ increase in
$f$ is sufficient to reduce the source density from (say)
$7/$bin to $1/(8\ {\rm bins})$. The ``unsubtractable part'' of
the galactic WD curve must therefore fall very steeply near $f \sim 1-2$
mHz---a feature accounted for by the exponential function in Eq.~(\ref{inst+gal}).
Also, a factor (say) $2$ change in our fiducial value for $\kappa T_{\rm mission}^{-1}$ would only
shift the ``drop-off'' frequency by a factor $2^{3/11} \approx  1.2$.
Therefore we believe Eq.~(\ref{inst+gal}) must
be a reasonably accurate representation of the total effective noise.

There are likely tens of thousands of CO capture sources
(out to $z=1$) in the LISA band at any instant, and we shall
see that a signficant fraction of them are not resolvable.
How should we modify Eq.~(\ref{inst+gal}) to take into account confusion
noise from CO captures? We shall see that even if {\it all} CO captures were
unsubtractable, so that the entire capture background is ``counted''
as confusion noise,
the CO confusion noise would still be smaller
than the (pre-subtraction) noise from GWDBs, though it
could be larger than the instrumental noise (depending on the exact
capture rates). In this case, it seems sensible, as a first approximation, to
simply add the CO confusion noise contribution, $S_h^{\rm capt}(f)$,
to $S_h^{\rm inst}(f)$ in Eq.~(\ref{inst+gal}). That is, we estimate the
total effective noise density as
\begin{eqnarray}\label{Stot}
S^{\rm eff}_h(f) &=& {\rm min}
\left\{  \left[S_h^{\rm inst}(f) + S_h^{\rm EGWDB}(f) + S_h^{\rm capt}(f)\right]
\exp(\kappa T^{-1} dN/df), \right. \nonumber\\
&& \quad\quad\quad \left.
S_h^{\rm inst}(f) + {\cal S}_h^{\rm GWDB}(f) + S_h^{\rm EGWDB}(f) +
S_h^{\rm capt}(f)\right\}.
\end{eqnarray}

Note that Eq.~(\ref{Stot}) does not include the effect of
``subtraction errors'' made in removing the resolvable CO captures.
We can reasonably neglect this noise contribution, for the following
reason.  Over a $\sim 3-$yr mission, the number of individually
detected captures will probably be $\lesssim 3\times 10^3$~\cite{emri}. Each
capture source is completely specified by $17$ parameters~\cite{BC}; thus $\sim
5 \times 10^4$ real numbers are required to specify their combined
signal. In comparison, the signal (from both synthetic detectors)
between $1$ and $10$ mHz is represented by $\sim 4 \times 10^6$ real
numbers (i.e., roughly $10^6$ discrete frequency bins, times two real
numbers per bin, times two independent detectors), so in principle
only $\sim 1\% $ of the available bandwidth is ``lost'' by fitting out
these captures.

\subsection{Noise spectral density}

\begin{figure}[htb]
\centerline{\epsfysize 10cm \epsfbox{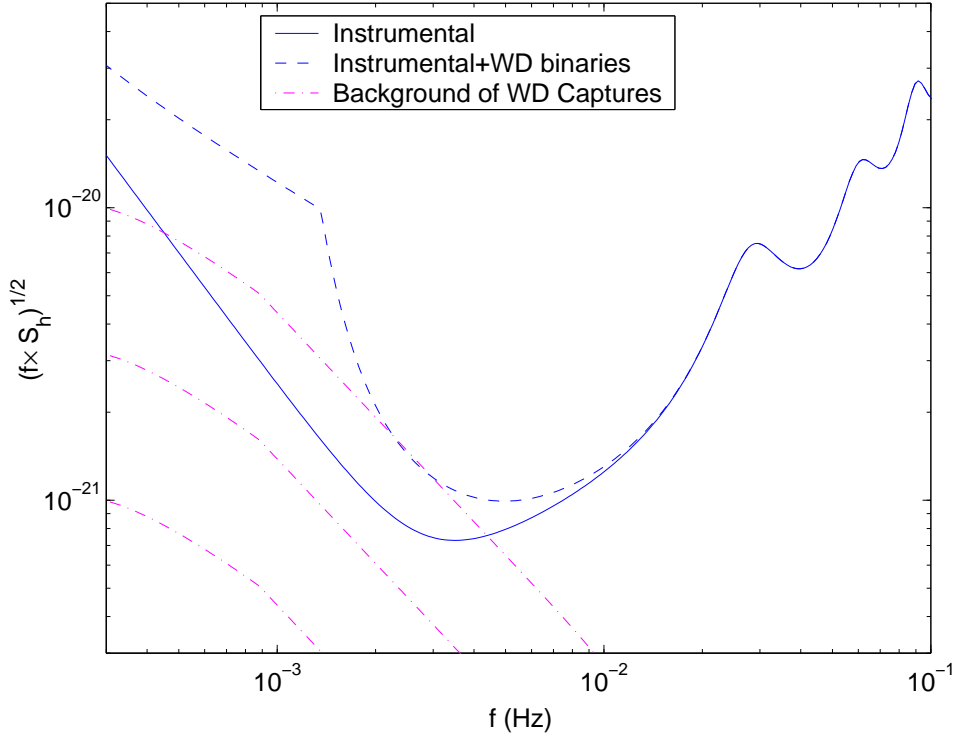}}
\caption{\protect\footnotesize
Comparison of the background ${\cal S}_h^{\rm WDcapt}$ from all WD captures
(dash-dot lines)
with LISA's instrumental noise (solid line) and with LISA's instrumental
plus confusion noise from WD binaries (dashed line). We show three cases,
corresponding to WD capture rates of $\kappa^{\rm WD} = 4\times 10^{-6},
4\times 10^{-7}$, and $4\times 10^{-8}$---cf.\ Eqs.\ (2) and (3).
}
\label{fig-SNR-WD}
\end{figure}

\begin{figure}[htb]
\centerline{\epsfysize 10cm \epsfbox{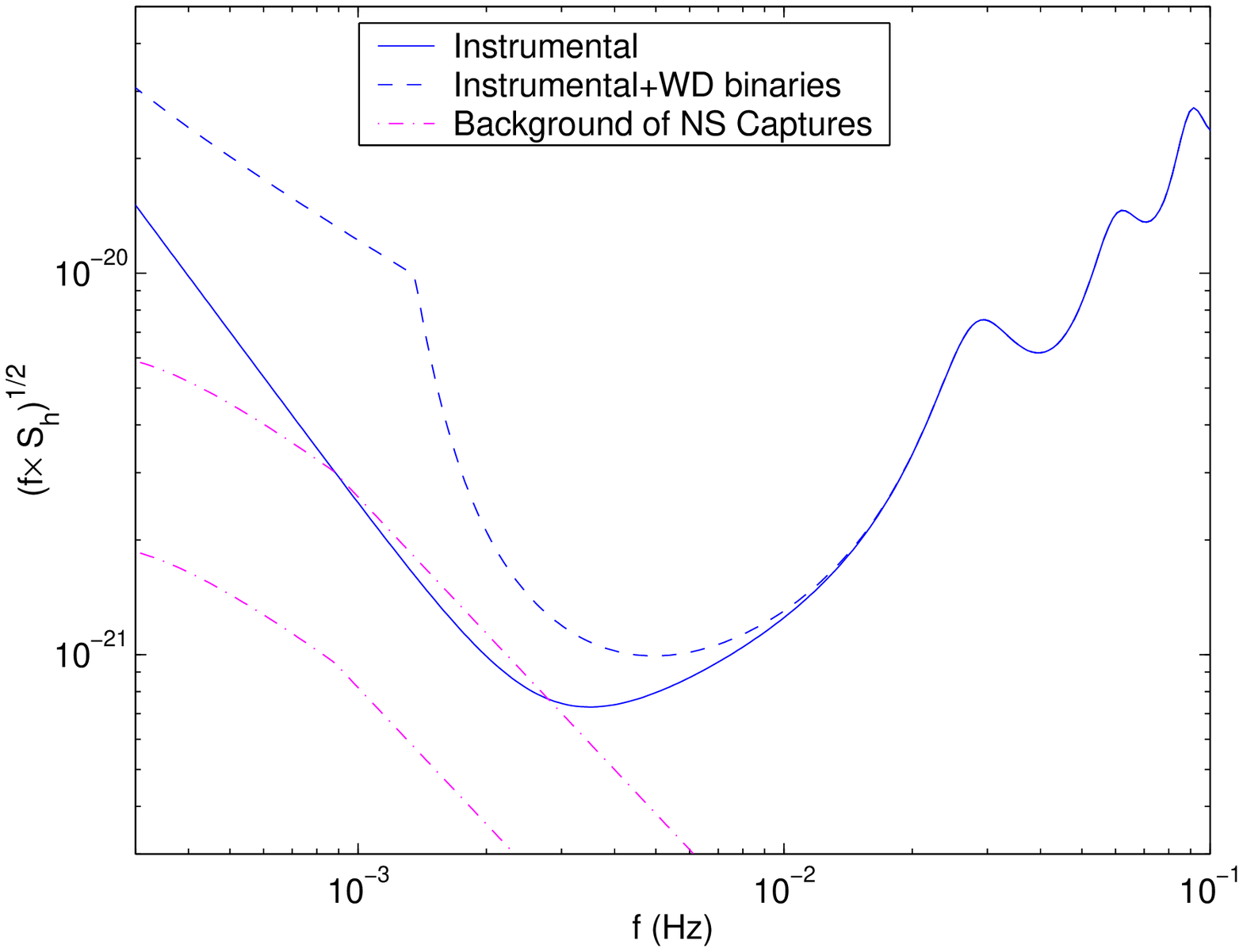}}
\caption{\protect\footnotesize
Same as in Fig.\ \ref{fig-SNR-WD}, but this time showing the background
${\cal S}_h^{\rm NScapt}$ from all NS captures.
We show two cases, corresponding to
NS capture rates of $\kappa^{\rm NS} = 6\times 10^{-7}$ and
$6\times 10^{-8}$.
}
\label{fig-SNR-NS}
\end{figure}

\begin{figure}[htb]
\centerline{\epsfysize 10cm \epsfbox{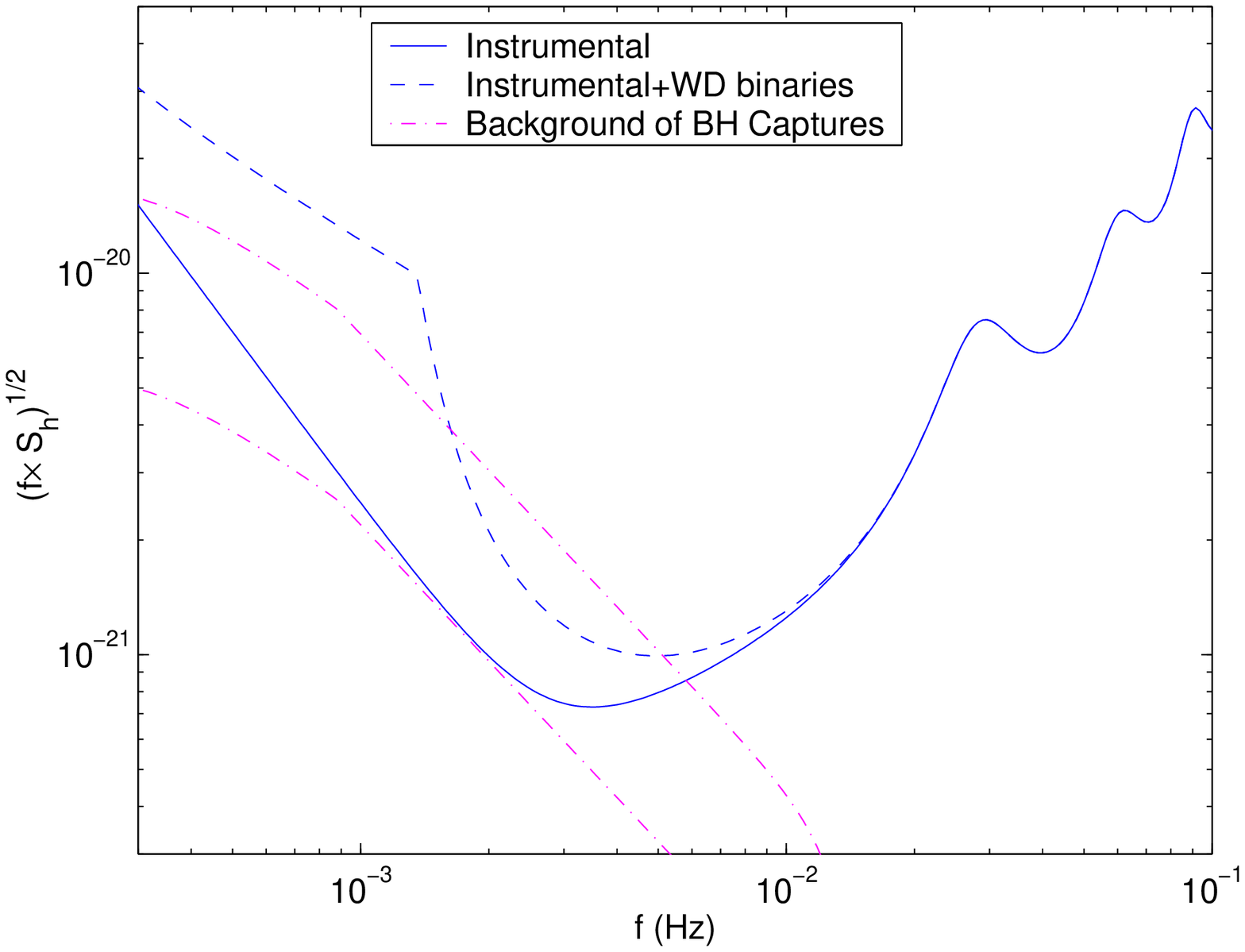}}
\caption{\protect\footnotesize
Same as in Fig.\ \ref{fig-SNR-WD}, but this time showing the background
${\cal S}_h^{\rm BHcapt}$ from all BH captures.
We show two cases, corresponding to
BH capture rates of $\kappa^{\rm BH} = 6\times 10^{-7}$ and
$6\times 10^{-8}$.
}
\label{fig-SNR-BH}
\end{figure}

We now obtain ${\cal S}^{\rm Acapt}_h(f)$---the spectrum of the background
from {\it all} capture sources
of type A---by simply combining Eqs.\ (\ref{epsilon}), (\ref{drhodf}),
and (\ref{ShOm}). This yields
\begin{eqnarray} \label{S}
{\cal S}^{\rm Acapt}_h(f)&=&
10^{-39} \ {\rm Hz}^{-1}\left(\frac{m^A}{M_{\odot}}\right)
\left(\frac{\alpha}{0.06}\right)\left(\frac{\beta}{0.7}\right)
\left(\frac{\kappa^{A}}{10^{-7}}\right)\,\gamma\, h_{70}^2
\nonumber\\
&&\times  \left\{
\begin{array}{ll}
18.540                   & f_{\rm mHz}<0.22,       \\
1.778 f^{-3}_{\rm mHz} (1-0.5039\,f_{\rm mHz}^{-3/8}),      & 0.22<f_{\rm mHz}<0.88, \\
0.799 f_{\rm mHz}^{-27/8} - 3.302 \times 10^{-6},
& 0.88<f_{\rm mHz}<22,\\
1.695 f^{-3}_{\rm mHz} (f_{\rm mHz}^{-3/8}-0.1866),       & 22<f_{\rm mHz}<88,\\
0,                               & f_{\rm mHz}>88.
\end{array}
\right.\
\end{eqnarray}
Recall that here $\alpha$ is the total amount of energy (per
CO's mass $m$) radiated over the course of a single inspiral (cf.\ Fig.\ \ref{fig-alphaAv}),
$\beta$ is the time (in units of $10^{10}$ yr) during which captures
occur, $\kappa^A$ is an event rate parameter [see Eq.\ (\ref{kappa})],
and $\gamma$ is a parameter of order unity that normalizes the
space-density of MBHs.
This capture background is plotted in Figs.\ \ref{fig-SNR-WD},
\ref{fig-SNR-NS},
and   \ref{fig-SNR-BH} (for captures of WDs, NSs, and BHs, respectively),
superposed on LISA's instrumental noise and WD-binaries confusion noise.
Each of these figures plots the background for a range of assumed
capture rates $\kappa^A$; the other parameters in Eq.~(\ref{S}) are all
assumed to have their fiducial values: $\alpha = 0.06$, $\beta = 0.7$,
and $\gamma = 1.0$.  The WD, NS, and BH masses are $0.6, 1.4$, and $10 M_{\odot}$, respectively.


\subsection{Influence of cosmology and source evolution}
\label{Cosmology}

Our calculation of ${\cal S}_h^{\rm capt}(f)$  in the previous subsection
ignored source evolution and treated
the spacetime as flat---counting all sources
within $\beta \times 10^{10}$ lt-yr of us and ignoring the rest.
Here we repeat the calculation but for a realistic cosmology and a few
different evolutionary scenarios. We shall see that if we
set $\beta$ to $0.7$, then
in the crucial $1-5\,$Hz range (where capture noise {\it could} possibly have
a significant impact on LISA's total effect noise level),
this more-accurate treatment results in differences from our flat
model at a level
$\lesssim 15 \% $. This uncertainty associated with
source evolution is still much smaller than the uncertainty in
the {\it current} event rate.

Let $\dot{\cal E}(f,z)\Delta f$ be the rate (per unit proper time,
per unit co-moving volume) at which
GW energy between frequencies
$f$ and  $f + \Delta f$ is emitted into the universe, at red-shift $z$.
(Here $f$ is the frequency measured
by a contemporaneous observer, not the red-shifted frequency we measure
today.)
For instance, for source type A, the rate today is
$\dot{\cal E}(f,0) = {\cal R}^A(0) m^A f\epsilon(f)$.
The dependence of $\dot{\cal E}(f,z)$ on $z$ arises principally from
two effects. Firstly, the capture rate in the past, ${\cal R}^A(M;z)$ (measured
per unit proper time, per MBH) can differ from today's rate ${\cal R}^A(M;0)$.
Secondly, the MBHs were smaller in the past (since they are growing
by accretion),
which implies that the spectrum $\dot{\cal E}(f,z=2)$ (say) is blue-shifted
compared to the spectrum $\dot{\cal E}(f,z=0)$. (We stress that this blue-shifting is a
source-evolution effect. There is also a cosmological red-shift, to be
accounted for momentarily.) The total number of MBHs (per unit comoving
volume) may also evolve due to mergers, but we expect this effect
to be small between $z\lesssim 2$ and now, and so for simplicity we take
this total number to be constant.

To calculate the GW energy density {\it today}, in some
frequency band of width $\Delta f$, one integrates the contributions from
all times, after appropriate red-shifting of the waves' frequency and
energy:
\be\label{Eq:drho1}
\frac{d \rho}{df}\Delta f = \int_{z_{\rm max}}^0
\frac{\dot{\cal E}[f(1+z),z]}{1+z} \Delta f' \frac{d\tau}{dz} dz,
\ee
\noindent
where $\Delta f' = \Delta f(1+z)$ is the blue-shifted (relative to today)
frequency range
of the GWs when initially emitted, and the factor $1+z$ in the
denominator 
accounts for
the decrease in the waves' energy due to the redshift between $z$ and now.
Obviously the two $(1+z)$ factors cancel, so Eq.~(\ref{Eq:drho1}) can be
simplied to~\cite{owenetal,phinney}
\be\label{Eq:drho2}
\frac{d \rho}{df} = \int_{z_{\rm max}}^0
\dot{\cal E}[f(1+z), z]\frac{d\tau}{dz} dz.
\ee

We shall assume a spatially flat $(\Omega = 1.0$)
Friedman-Lema$\hat{\rm i}$tre-Robertson-Walker
universe, with $\Omega_\Lambda = 0.70$ and $\Omega_m = 0.30$.
For our fiducial cosmology, the universe's current age $\tau_0$ is given by
\be
\tau_0 = 0.964 H_0^{-1} = 1.39 \times 10^{10} h_{70}^{-1}\ {\rm yr}.
\ee
Here we shall consider only sources
in the range $0 < z <2$. (We will see below that sources at $z>2$ are unlikely
to add considerably to $\rho_{\rm GW}$.)
In this range, the following power law
relation between the universe's age $\tau$ and redshift $z$ is accurate
to within $\sim 3\%$:
\be
\tau = \tau_0 (1+z)^{-1.18}\, .
\ee
For our estimates in this section, we will approximate this as
$\tau = \tau_0 (1+z)^{-1.2}$.

To get some idea for the range of possible answers, we shall consider two
possible scalings for the capture rates ${\cal R}^A(z)$ and two
possible scalings for the MBH mass $M(z)$ (i.e., four possible scenarios in
all):
For the capture rates,
simulations of Milky Way captures by Freitag \cite{Freitag}, as well as
an analytic argument by Phinney \cite{PhinneyUP}, both suggest
that ${\cal R}^A(z)$ should increase in the past like
$\tau^{-1/2}\propto (1+z)^{0.6}$. For comparison, we will also consider
an ${\cal R}^A(z)={\rm const}$ scenario.
Similarly, for the MBH masses, we will consider as possibilities
(i) $M = {\rm const}$
and (ii) $M \propto \tau^{1/2}$ (as suggested by Sirota {\it et al.~}\cite{Sirota}).
The four different assumptions yield the following four different
relations between $ \dot{\cal E}(f,z)$ and $ \dot{\cal E}(f,0)$:

\begin{eqnarray} \label{Ez}
\dot{\cal E}(f,z) &=&
\left\{
\begin{array}{ll}
 \dot{\cal E}(f,0) ,               & {\cal R}^A(z) = {\rm const},\, M = {\rm const} ,      \\
(1+z)^{0.6}\, \dot{\cal E}(f,0) ,      &   {\cal R}^A(z) \propto (1+z)^{0.6},\, M = {\rm const} ,   \\
(1+z)^{-0.6}\, \dot{\cal E}[f(1+z)^{-0.6},0] , & {\cal R}^A(z) = {\rm const},\, M \propto (1+z)^{-0.6} , \\
 \dot{\cal E}[f(1+z)^{-0.6},0]      &   {\cal R}^A(z) \propto  (1+z)^{0.6},\, M \propto (1+z)^{-0.6} .
\end{array}
\right.\
\end{eqnarray}
To derive the third line on the RHS, note that under a rescaling of MBH mass, $M \rightarrow
M^\prime = \lambda M$, with total luminosity fixed, the spectrum gets re-scaled
according to $\dot{\cal E}^\prime(f^\prime) \Delta f^\prime =
\dot{\cal E}(f)\Delta f $, where $f^\prime = \lambda^{-1} f$, or
\be
\dot{\cal E}^\prime(f^\prime) = \lambda \, \dot{\cal E}(\lambda f^\prime) \, .
\ee
In our case, $M^\prime$ is the MBH mass at redshift $z$, so $\lambda = (1+z)^{-0.6}$, and
the third line follows when we replace the dummy variable  $f^\prime$ by $f$.
Lines 2 and 4 on the RHS of Eq.~(\ref{Ez}) are obtained by simply multiplying lines 1 and 3
(respectively) by $(1+z)^{0.6}$ (the assumed rate enhancement factor for early times).

Again, we
are interested in the CO capture spectrum chiefly in the
range $1\lesssim f\lesssim 7$ mHz.
If we are to integrate Eq.~(\ref{Eq:drho2})
out to $z=2$
then we must know $\dot{\cal E}(f,0)$
for $f$ between $1$ mHz and $7\times (1+2) = 21$ mHz.
In this range, $\dot{\cal E}(f,0)$ is basically a power law:
$\dot{\cal E}(f,0)\propto f^{-1}\epsilon(f) \propto f^{-11/8} $ (see Fig.~5).

For the remainder of this section, we will approximate $\dot{\cal E}(f,0)$ by
the above simple power law; i.e., we assume $\dot{\cal E}(f,0) =   W\,  f^{-11/8}$ for some
$W$. 
Using $d\tau/dz = -1.2 \tau_0 (1+z)^{-2.2}$
and integrating Eq.~(\ref{Eq:drho2}) out to $z=2$, we then obtain
\be\label{Eq:drho4}
\frac{d \rho}{df} = 1.2\,  \tau_0\,  W\, f^{-11/8}
\int_1^3{ y^{-3.575 + \mu +\sigma}dy},
\ee
where $y \equiv (1+z)$.
In the exponent, $\mu = 0$ [for $R^A(z) =  R^A(0)$]
or $0.6$  [for $R^A(z) =  (1+z)^{0.6} R^A(0)$], and
$\sigma = 0$ [for $M(z) =  M(0)$]
or $(-0.6)(-11/8 +1) = 0.225$  [for $M(z) = M(0)(1+z)^{-0.6}$].
Note that the shape of the spectrum is always the same; only the
amplitude varies with evolutionary scenario.
We can write the result in the form
\be
\frac{d \rho}{df} = W\,\tau_{\rm eff}\, f^{-11/8},
\ee
for some ``effective'' integration time $\tau_{\rm eff}$.
Performing the elementary integral in Eq.~(\ref{Eq:drho4}) (taking
$h_{70}=1$), we find for our 4 cases:
$\tau_{\rm eff} = 6.1 \times 10^{9}$yr ($\mu=\sigma=0$),
 $\tau_{\rm eff} = 7.5 \times 10^{9}$yr ($\mu=0.6, \sigma=0$),
$\tau_{\rm eff} = 6.6 \times 10^{9}$yr ($\mu=0, \sigma= 0.225$),
and $\tau_{\rm eff} = 8.1 \times 10^{9}$yr ($\mu=0.6, \sigma= 0.225$).

By comparison, $\tau_{\rm eff}$ for the flat-universe/no-evolution model considered in
the previous subsection would just be the total integration time: $\beta\times 10^{10}$yr.
Thus, if we make the choice $\beta = 0.7$, our flat-universe/no-evolution
calculation agrees with all four of our cosmological/evolutionary
models to within $\sim 15 \%$. We note that $7 \times 10^{9}$ years ago
corresponds to $z  = 0.79$.
We also note that in all four models, most of the contribution to the integral
comes from the range $0< z < 1$; the contribution from $1< z < 2$
accounts for only  $12-18\%$ of the total (depending on the case).
This justifies our cutting off the integral at $z= 2$, which is
convenient, since our simple evolutionary models would not be trustworthy
at higher $z$.

\section{Confusion noise from captures}  \label{SecV}

In Sec.\ \ref{SecIV} we estimated ${\cal S}^{\rm capt}_h(f)$, the
spectrum of the background due to {\it all} captures. At
$f = 3$ mHz (i.e., near the bottom of LISA noise curve)
${\cal S}^{\rm capt}_h(f)$ {\it equals} the instrumental noise
level $S^{\rm inst}_h(f)$ for $\kappa^{\rm WD} = 1.6\times 10^{-6}$,
$\kappa^{\rm NS} = 6.8\times 10^{-7}$, or
$\kappa^{\rm BH} = 9.5\times 10^{-8}$.
At the high end of their estimated ranges, the actual rates are larger
than these values by factors of $\sim 2.5$, $1$, and $6$, respectively.
Thus, confusion noise from capture sources {\it could} have
a significant effect on the total LISA noise level, making
it is important to next consider what fraction of ${\cal S}^{\rm capt}_h(f)$
actually constitutes an unresolvable confusion background (or, equivalently,
what portion is resolvable and hence subtractable).
With this goal in mind, we first
make some general remarks on how the confusion noise level depends
on both the event rate and the available subtraction method.

\subsection{General scalings of source rate, detection rate, and confusion
noise} \label{Sec:scaling}

\begin{figure}[htb]
\centerline{\epsfysize 6cm \epsfbox{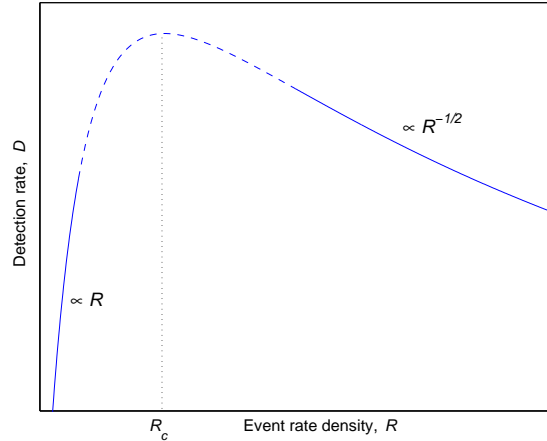}}
\caption{\protect\footnotesize
Illustration of how the detection rate $\cal D$ for some source class
$\cal C$ scales with the event rate $\cal R$ (number of sources
per unit time per unit space volume). For sufficiently small $\cal R$,
we have ${\cal D}\propto {\cal R}$, whereas for very large $\cal R$ we have
${\cal D}\propto {\cal R}^{-1/2}$.
Hence, the detection rate must peak at a certain event rate, ${\cal R}_c$.
${\cal R}_c$ is roughly the rate where confusion noise from
$\cal C$-type sources begins to dominate the total noise level.
}
\label{fig-DvsR}
\end{figure}

In this subsection, we step back and discuss the general
phenomenon of confusion noise in GW searches. Consider some class $\cal C$
of astronomical sources, having some
event rate $\cal R$, measured in ${\rm Mpc^{-3}yr^{-1}}$.
(For sources that are always ``on'', e.g.\ GWDBs,
we would define $\cal R$ as just the spatial density, measured in
${\rm Mpc^{-3}}$.)
We first ask:
How does the detection rate $\cal D$ vary with $\cal R$?
Here, for simplicity, we will imagine that source types {\it other} than
$\cal C$ contribute negligibly to confusion noise; i.e.,
all noise is either instrumental noise or confusion noise from
$\cal C$-type sources.
Then for sufficiently low $\cal R$, confusion noise will be
negligible, so the average distance $d_{\rm det}$ out to which
one can detect individual  $\cal C$-type sources is
a constant (i.e., independent of $\cal R$) set by the
instrumental noise level. Clearly, then, the detection rate $\cal D$
clearly grows linearly with $\cal R$ at small $\cal R$.
But as one increases $\cal R$,
eventually one must reach a point where confusion noise from
$\cal C$-type sources dominates over the instrumental noise, and so
determines $d_{\rm det}$.
Clearly, for sufficiently large $\cal R$,
almost all events add to the confusion noise, so $S_h^{\cal C}(f)
\propto {\cal R}$, and  therefore $d_{\rm det} \propto {\cal R}^{-1/2}$.
If $d_{\rm det}$ is in the regime where space can be
approximated as Euclidean and the source distribution can
be approximated as spherically symmetric (generally true when
$d_{\rm det}$ is much larger than typical
separations between galaxies, but much smaller than a Hubble length: roughly
$10 {\rm Mpc} < d_{\rm det} < 1 {\rm Gpc}$), then the detection rate ${\cal D} \propto
{\cal R} d_{\rm det}^3 \propto {\cal R}^{-1/2}$, for large ${\cal R}$.
 Thus the individual-source
detection rate $\cal D$ must peak at a certain rate ${\cal R}_c$, which
is roughly the event rate at which confusion noise starts to dominate over
instrumental noise in setting $d_{\rm det}$.
This is illustrated in Fig.\ \ref{fig-DvsR}.

\begin{figure}[htb]
\centerline{\epsfysize 6cm \epsfbox{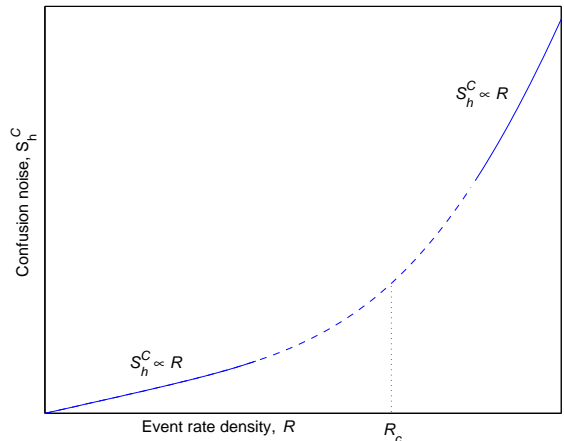}}
\caption{\protect\footnotesize
Illustration of how the confusion noise $S_h^{\cal C}$ scales with
the event rate density $\cal R$. $S_h^{\cal C}$ is linear in $\cal R$
in both the very small $\cal R$ and very large $\cal R$ regimes.
It depends on $\cal R$ non-linearly around the rate ${\cal R}_c$, where
confusion noise begins to dominate the total noise level.}
\label{fig-SvsR}
\end{figure}

Next, we ask how the
confusion noise $S_h^{\cal C}(f)$ from unresolvable sources scales with
the rate ${\cal R}$.\footnote{Of course,
$S_h^{\cal C}(f)$ is a function, not a single number, so here we mean
its value at some typical frequency, where it strongly affects, and is
affected by, the detection rate ${\cal D}$.}
For low values of ${\cal R}$,
$S_h^{\cal C}(f)$ should be a fixed fraction of ${\cal S}_h^{\cal C}(f)$
(arising from those sources whose SNR is too small to permit individual
detection), and so grows linearly with $\cal R$.
On the other hand, for very high ${\cal R}$, where confusion noise limits the detection rate,
the fraction of source energy that is ``unresolvable'' grows and
approaches one at sufficiently high ${\cal R}$; i.e.,
$S_h^{\cal C}(f)$ approaches ${\cal S}_h^{\cal C}(f)$.
Thus,  $S_h^{\cal C}(f)$
grows approximately linearly with ${\cal R}$ at very high ${\cal R}$, too,
but with
much larger slope than at low ${\cal R}$.
Between these two regimes, near ${\cal R}_c$,   $S_h^{\cal C}(f)$ grows
nonlinearly with  ${\cal R}$---see Fig.\ \ref{fig-SvsR}.

To recapitulate: First, we have observed that the event
rate ${\cal R}$ where ${\cal S}_h^{\cal C}$
begins to dominate the total noise
is roughly the event rate that gives
the highest individual-source detection rate
for ${\cal C}$-type sources---so one is  lucky to be in that
position! Second, the confusion noise level $S_h^{\cal C}$
is generally hardest to estimate precisely when it begins to dominate, since
there it depends nonlinearly on ${\cal R}$.
It is easy to see the problems that arise when
$S_h^{\cal C}(f)$ vs. ${\cal R}$ is in the nonlinear regime:
When confusion noise is the dominant noise source,
one cannot determine which sources are detectable without first knowing
the confusion noise
level; yet, to calculate the confusion noise level, one must
know which sources are detectable and hence subtractable.

We imagine that in practice one gets around this ``chicken-and-egg problem''
by an iterative procedure of
the following sort.  At the very beginning, before any sources
have been identified, the entire background must be treated as confusion noise.
Then one identifies the very brightest sources (i.e., those with highest SNR)
and subtracts them. With the noise level so reduced, it will be possible
to identify somewhat weaker sources. One subtracts them as well, and iterates
until there is no clear improvement. The confusion noise is roughly ``what's left
over'' from the background after the detectable sources have all been
subtracted. (Note that in practice, for LISA, this strategy is complicated
by the fact that there are at least two distinct and important
sources of confusion noise: CO captures and compact binaries.  Moreover,
the galactic binaries are always ``on'', and so
as time goes by, more and more galactic binaries will be
individually identified, and then subtracted from the {\it entire} LISA
data stream. Thus a CO capture that plunges in 2014, and is slightly
too faint to be detected then, might become ``visible'' in the old data
starting in 2017, thanks to improved cleaning out of galactic binaries.)

Finally, we mention that there are cases where
$S_h^{\cal C}(f)$ is particularly simple to estimate:
If only a tiny percentage of ${\cal C}$-type sources can be
individually  identified even at very
small ${\cal R}$ (i.e., even at rates where $S_h^{\cal C}(f)$
does {\it not} dominate the total noise), then
$S_h^{\cal C}(f)$ is approximately ${\cal S}_h^{\cal C}(f)$ for {\it all}
rates ${\cal R}$. We shall see that  WD and NS captures are basically
in this ``easy to estimate'' category, but captures of $10 M_{\odot}$ BHs are not.
For the latter, we shall content ourselves with some less-accurate
estimates (rather than trying now to simulate the iterative procedure outlined
above).

\subsection{Dependence on subtraction method}

We want to estimate what fraction of ${\cal S}_h^{\rm Acapt}$ is due to
detectable sources.
Unfortunately, computationally practical methods for detecting captures
are likely to be substantially less sensitive than the ``optimal''
(but here completely impractical) method of coherent matched filtering.
The parameter space of CO capture waveforms is 17-dimensional;
a full matched-filter search for a {\it single} source over this space
(using a discrete set of template that densely cover the space like a mesh),
has been estimated (very roughly) to require
$\sim 10^{40}$ templates~\cite{emri}.
Therefore, the threshold signal-to-noise ratio
${\rm SNR}_{\rm thresh}$ required for a $1 \%$ false alarm probability
in a search over the entire template bank
is given by ${\rm erf}({\rm SNR}_{\rm thresh}/\sqrt{2})
= 10^{-42}$,
or ${\rm SNR}_{\rm thresh} \approx 14$. (Note that at such high thresholds,
the exact value of the threshold is quite insensitive to
our estimate of the number of required templates; increasing the estimate
to $10^{50}$ templates only raises the threshold to
${\rm SNR}_{\rm thresh} \approx 15$,
assuming Gaussian noise.)

Here SNR means the optimal, matched
filtering SNR using the complete LISA data set---approximated in this paper
as the output of two orthogonal Michelsons.
There are some well known tricks for speeding up the search (e.g.,
of the $\sim 10^{40}$ templates mentioned above, a factor $\sim 10^{5}$
comes from simple time translations of otherwise identical templates, and
this subspace of time-translations can be searched very efficiently using
FFTs~\cite{schutz_book}).
Still, it is completely
impractical to imagine searching directly over this vast set of
templates.
In Gair {\it et al.}\ \cite{emri}, a strategy
was outlined of searching for CO captures in a hierarchical fashion; the
signal strength required for detection with this strategy was estimated
to be ${\rm SNR}_{\rm thresh} \approx 30$, for a 3-yr integration with
realistic computing power (where ``realistic computing power'' at the start
of the LISA mission in $\sim 2013$ was estimated
using Moore's Law).

  For the purpose of estimating confusion noise levels, in this paper
we shall assume that individual sources are detectable in practice if
their 3-yr, matched-filtering ${\rm SNR}$ (in detectors I and II combined)
is $\gtrsim {\rm SNR}_{\rm thresh}$,  where ${\rm SNR}_{\rm thresh} = 30$.
We shall also see that our basic conclusions would change very little
if the ${\rm SNR}_{\rm thresh}$ were half or double this value.


\subsection{Estimating the fraction of subtractable capture sources}

We now estimate what fraction of the capture background
${\cal S}_h^{\rm Acapt}(f)$
is subtractable, and what is the leftover
portion that constitutes the capture confusion noise $S_h^{\rm Acapt}(f)$.
We estimate this as follows.
Let  $\rho^{\rm Acapt}_{\rm GW}$ be the local (i.e., near LISA) GW energy density
from captures of type A (again, A = WD, NS, or BH).
We ask what fraction $U^{\rm A}$ of $\rho^{\rm Acapt}_{\rm GW}$ is due
to sources that are undetectable by LISA. We then estimate that
$S_h^{\rm Acapt}(f) = U^{\rm A} {\cal S}_h^{\rm Acapt}(f)$.
Of course, this estimate is quite crude---in particular, it ignores
any frequency-dependence in the ratio
$S_h^{\rm Acapt}(f)/{\cal S}_h^{\rm Acapt}(f)$---but it seems good
enough for our purposes.
[A less crude version would be to subdivide A into many smaller subclasses,
parametrized by (at least) the MBH mass and the time $\tau$ left before
final plunge. One would then estimate the spectrum from
undetectable sources in each subclass, and then average all the spectra
(weighted appropriately) to get $S_h^{\rm Acapt}(f)$.
This would give a nontrivial frequency dependence to
$S_h^{\rm Acapt}(f)/{\cal S}_h^{\rm Acapt}(f)$.]

\begin{figure}[htb]
\centerline{\epsfysize 6cm \epsfbox{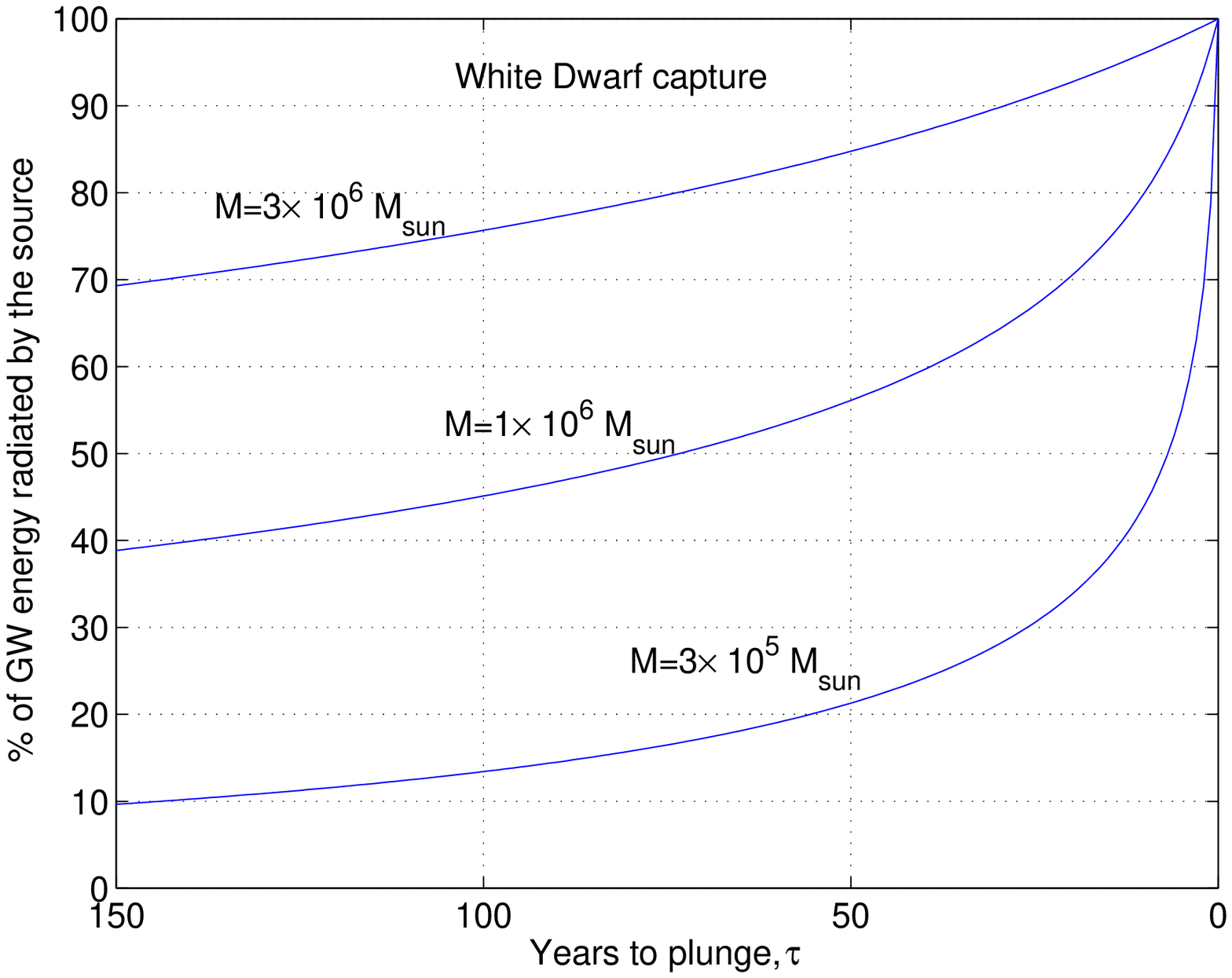}\hspace{4mm}
\epsfysize 6cm \epsfbox{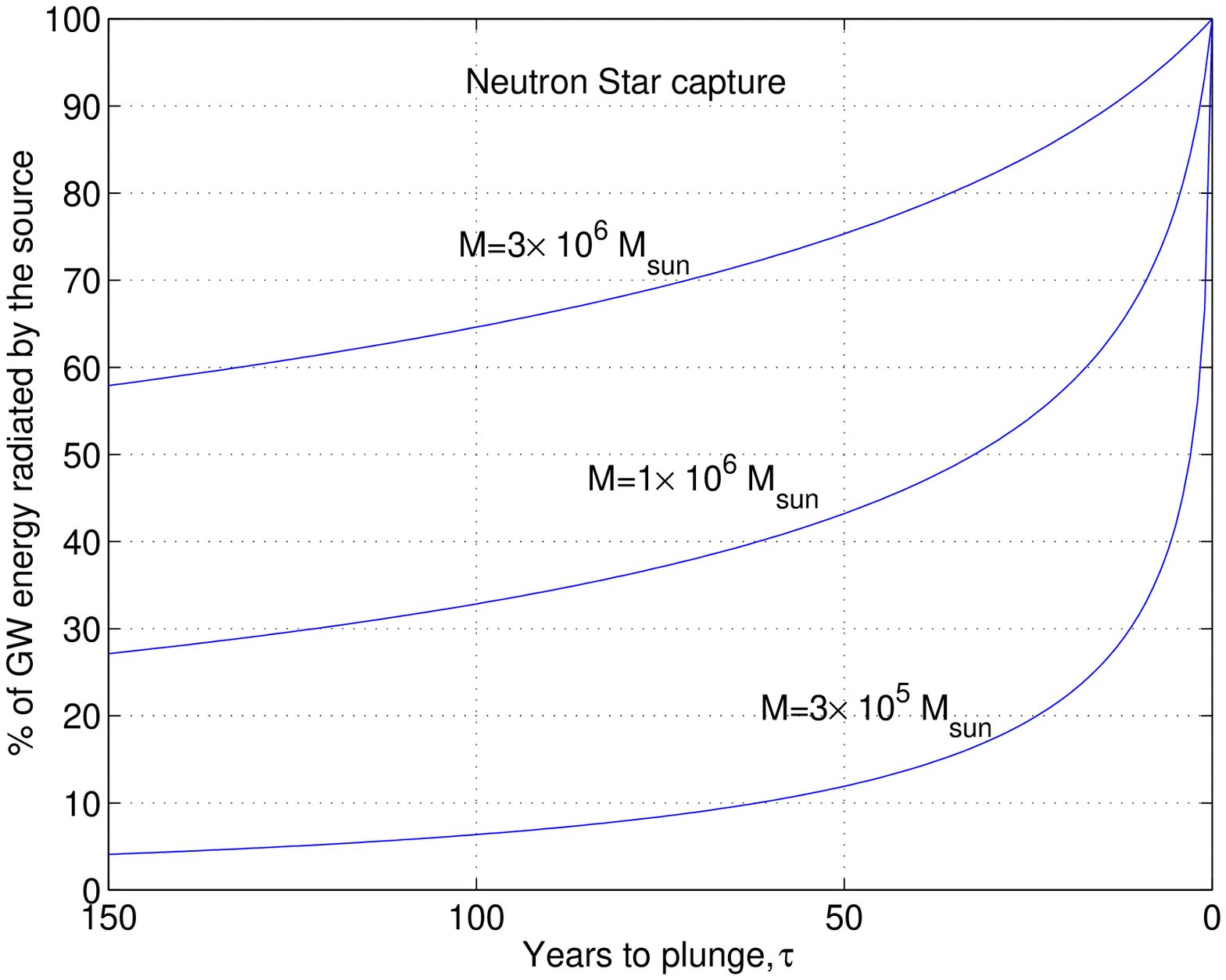}}
\centerline{\epsfysize 6cm \epsfbox{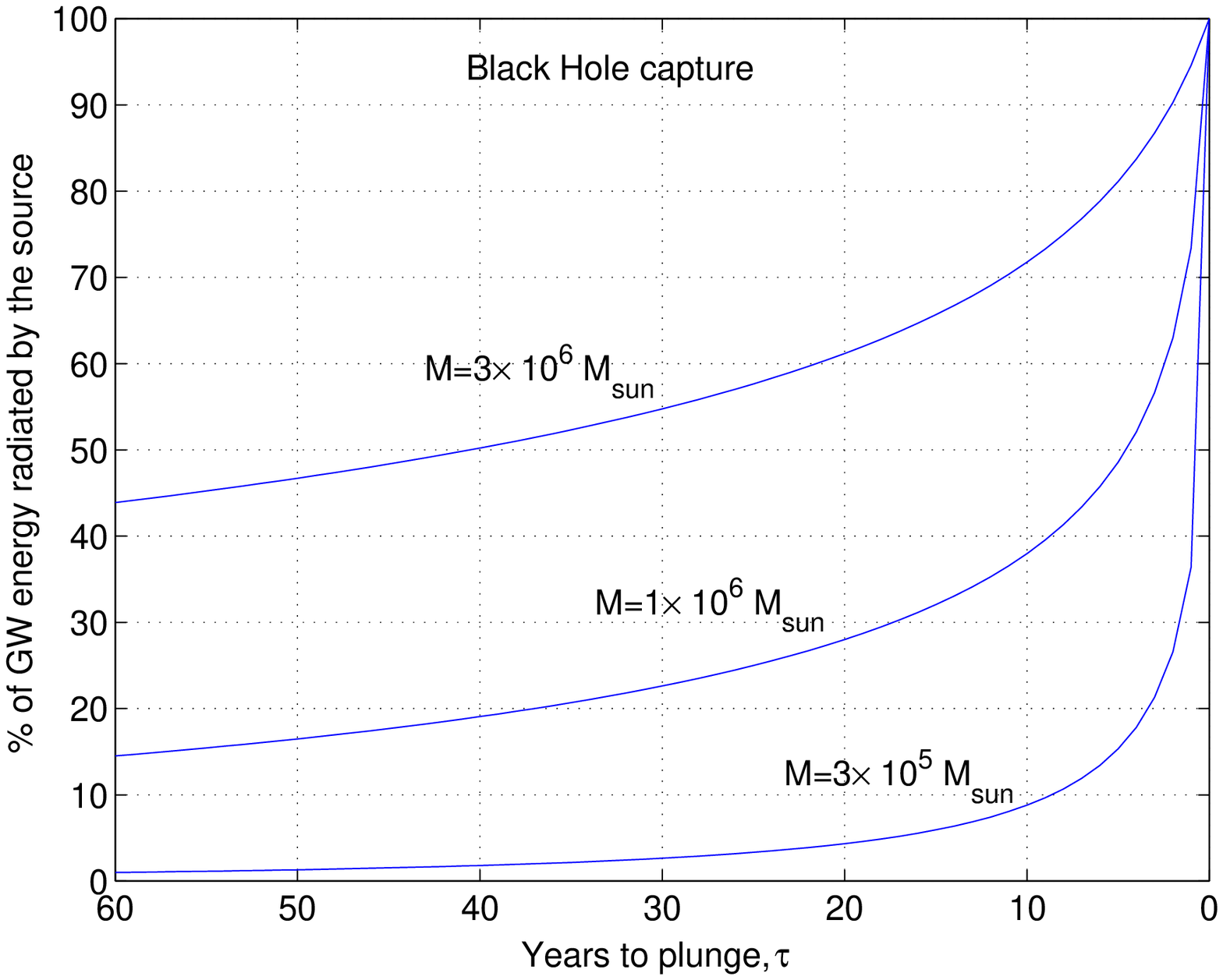}}
\caption{\protect\footnotesize
Percentage of GW energy radiated (out of the total energy radiated
during the entire inspiral) as a function of the time $\tau $ left to plunge, for
typical captures of a WD (upper left panel), a NS (upper right panel), and a
BH (lower panel), for a range of MBH masses.
The masses of the WD, NS, and BH are taken to be
$m=0.6,1.4$, and $10 M_{\odot}$, respectively, and in all cases the CO plunges
at eccentricity $e_{\rm LSO}=0.15$.
A substantial fraction of the energy is emitted long before plunge,
when the source is still not detectable by LISA. This energy is mostly emitted
in relatively high-frequency bursts near periastron passage, and
so contributes to
the confusion noise.
}
\label{fig-Erad}
\end{figure}

For our estimates, we will need to know how the GW luminosity
from capture sources varies over their inspiral history and
also how their detectability (i.e., their 3-yr SNR) increases over this time.
For a given capture, let $\tau$ be the time
left until the CO plunges into the MBH. We estimate the luminosity as a
function of time, $\dot{E}(\tau)$, using the
Peters and Matthews result~\cite{PM} (derived using the
quadrupole formula) for the quasi-Newtonian elliptical orbits:
\be\label{pm1}
\frac{dE}{dt} =  \frac{32}{5} (m/M)^2 (2\pi M\nu)^{10/3}(1-e^2)^{-7/2}
\left[1+(73/24)e^2+(37/96)e^4\right] .
\ee
\noindent Here $\nu$ and $e$ are the orbital frequency and eccentricity,
respectively, whose evolution is described to lowest order by~\cite{PM}
\begin{eqnarray}
\frac{d\nu}{dt} &=&
\frac{96}{10\pi}(m /M^3)(2\pi M\nu)^{11/3}(1-e^2)^{-7/2}
\left[1+(73/24)e^2+(37/96)e^4\right] \label{pneqs1} , \\
\frac{de}{dt}  &=& -\frac{e}{15}(m/M^2)(2\pi M\nu)^{8/3} (1-e^2)^{-5/2}
(304+121e^2) \label{pneqs2} .
\end{eqnarray}
\noindent We integrate Eqs.~(\ref{pneqs1}) and (\ref{pneqs2}) backwards
in time, from $\nu_{\rm LSO} =  (2\pi M)^{-1}[(1-e^2)/(6 + 2e)]^{3/2}$, to
obtain $\nu(t)$ and $e(t)$ [c.f. the discussion around Eq.~(59) in \cite{BC}].
We then integrate $\dot{E}(t)$ numerically to determine the
amount of energy $E(\tau)$ emitted before time $t= -\tau$, as well as the
fraction $E(\tau)/E(0)$, where $E(0)$ is the total GW energy
emitted from the beginning of the inspiral until plunge.
The outcome from this calculation is presented in Fig.\ \ref{fig-Erad}
for ``typical'' captures of a WD, a NS, and a BH (with $m=0.6, 1.4$, and $10 M_{\odot}$, respectively);
more specifically, we took $e_{\rm LSO} = 0.15$ and $a=0$ (i.e., the MBH is Schwarzschild),
and considered a range of MBH masses.
Note that a large fraction of the GW energy is emitted
long before the source is detectable by LISA. In a typical BH capture,
roughly $40\% $ the GW energy will have been emitted
already 10 years before the BH plunges. If the captured object is a WD,
$\sim 40\% $ of the energy will have been emitted 150 years before plunge.

\begin{figure}[htb]
\centerline{\epsfysize 8cm \epsfbox{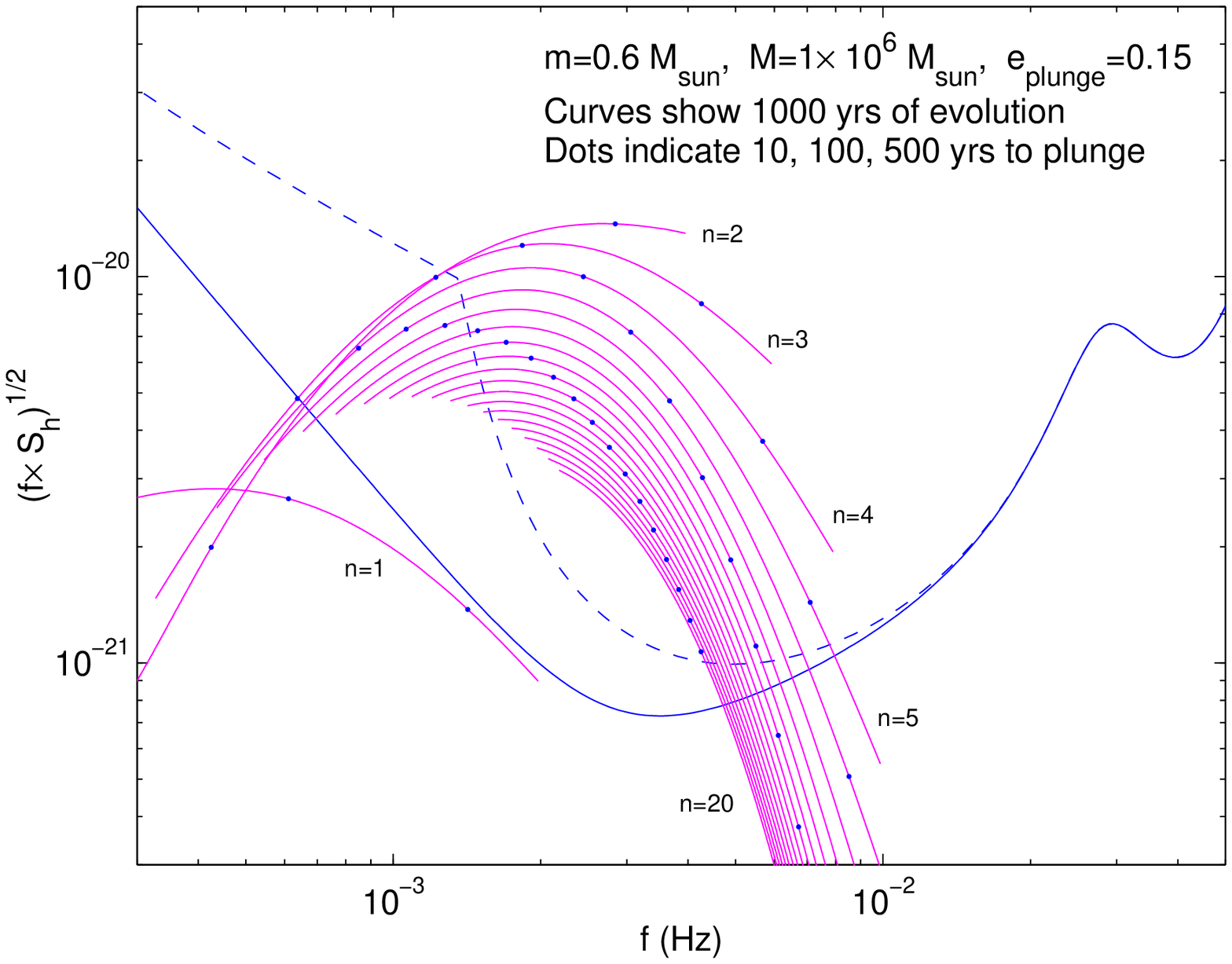}}
\caption{\protect\footnotesize
GW output from the last 1000 years of inspiral of a captured WD at 1 Gpc.
Shown (solid convex curves) are the characteristic amplitudes $h_{c,n}$
[see Eq.\ \ref{hcn}]
of each of the first 20 harmonics (labeled $n=1,\ldots, 20$), for
a capture of a $0.6 M_{\odot}$ WD by a $10^6 M_{\odot}$ nonspinning MBH,
with plunge eccentricity 0.15. The signal curves are superposed
on LISA's instrumental noise
(solid concave curve) and LISA's noise curve including confusion from WD
binaries (dashed curve). In each $h_{c,n}$ curve, the right end
corresponds to the instant of plunge,
the left end corresponds to 1000 years before plunge, and dots along the curve
correspond (from right to left) to 10, 100, and 500 years before plunge.
Note that even 500 years before plunge, most of the GW energy is radiated
``in band''---i.e., in the frequency range
where LISA is most sensitive.
Although at such early times the orbital frequency
is well ``below band," most of the energy radiated then is
carried by high harmonics of the orbital frequency.
}
\label{fig-WDsource}
\end{figure}

For such typical sources, the GWs emitted near plunge are in the
LISA band, so one
might naively have thought that the GWs emitted $10-150$ yr
before plunge would be at
frequencies well below the LISA band. But this is not true, since most of
the radiation is emitted in relatively high-frequency bursts (i.e., high
harmonics of the orbital frequency) near periastron passage.
To illustrate this, in Fig.\ \ref{fig-WDsource} we plot the signal from
a WD captured by a $10^6 M_{\odot}$ MBH at 1 Gpc.
This plot shows the evolution of the signal during the last 1000 years of
inspiral, as distributed among the first 20 harmonics of the radiation.
We see that early in the inspiral history, the radiation is dominated by
the high harmonics,
which are well within LISA's sensitivity band.
Thus the capture source is
effectively ``in band'' throughout the entire inspiral, and
so is always a potential source of confusion.

\begin{figure}[htb]
\centerline{\epsfysize 6cm \epsfbox{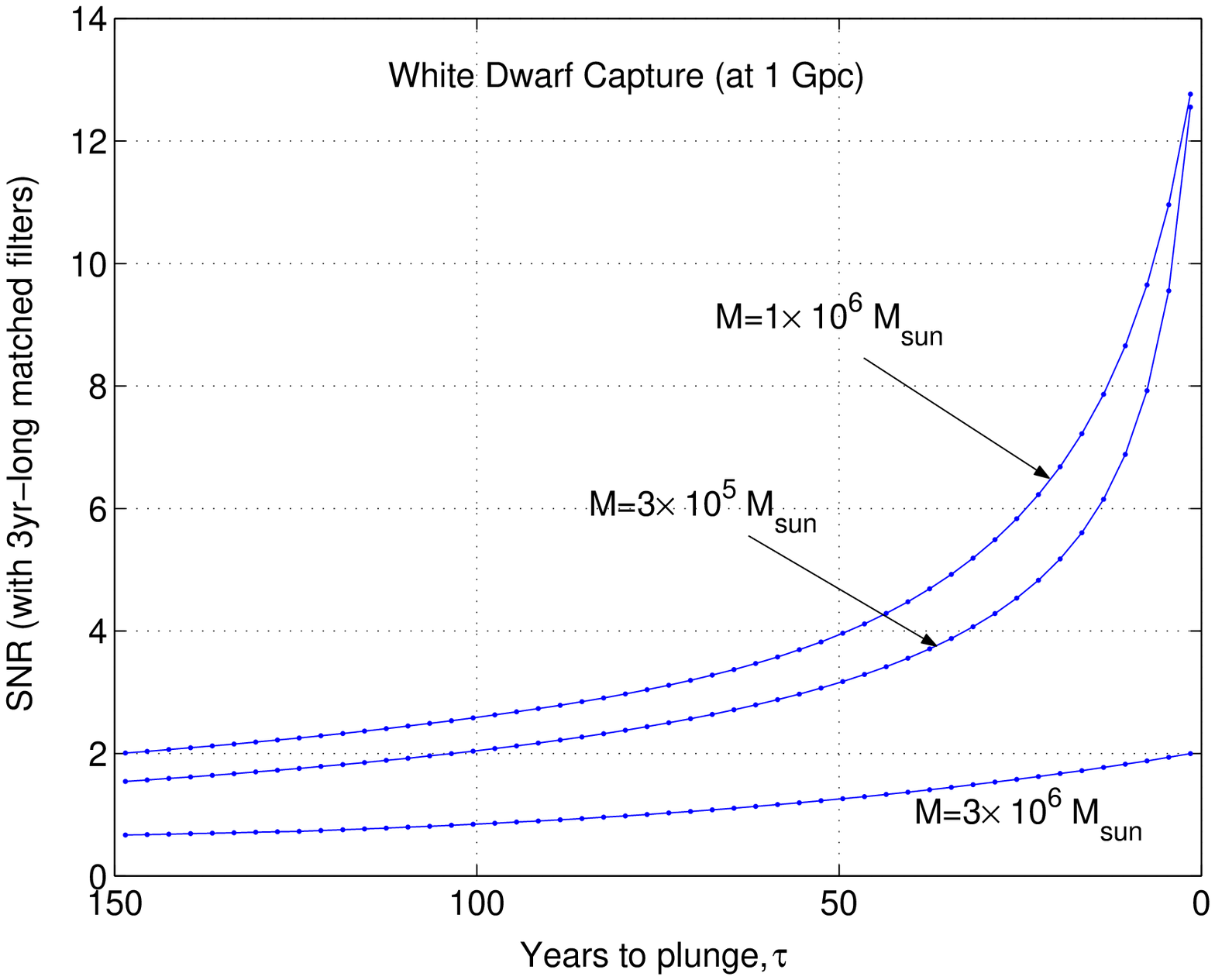}\hspace{4mm}
\epsfysize 6cm \epsfbox{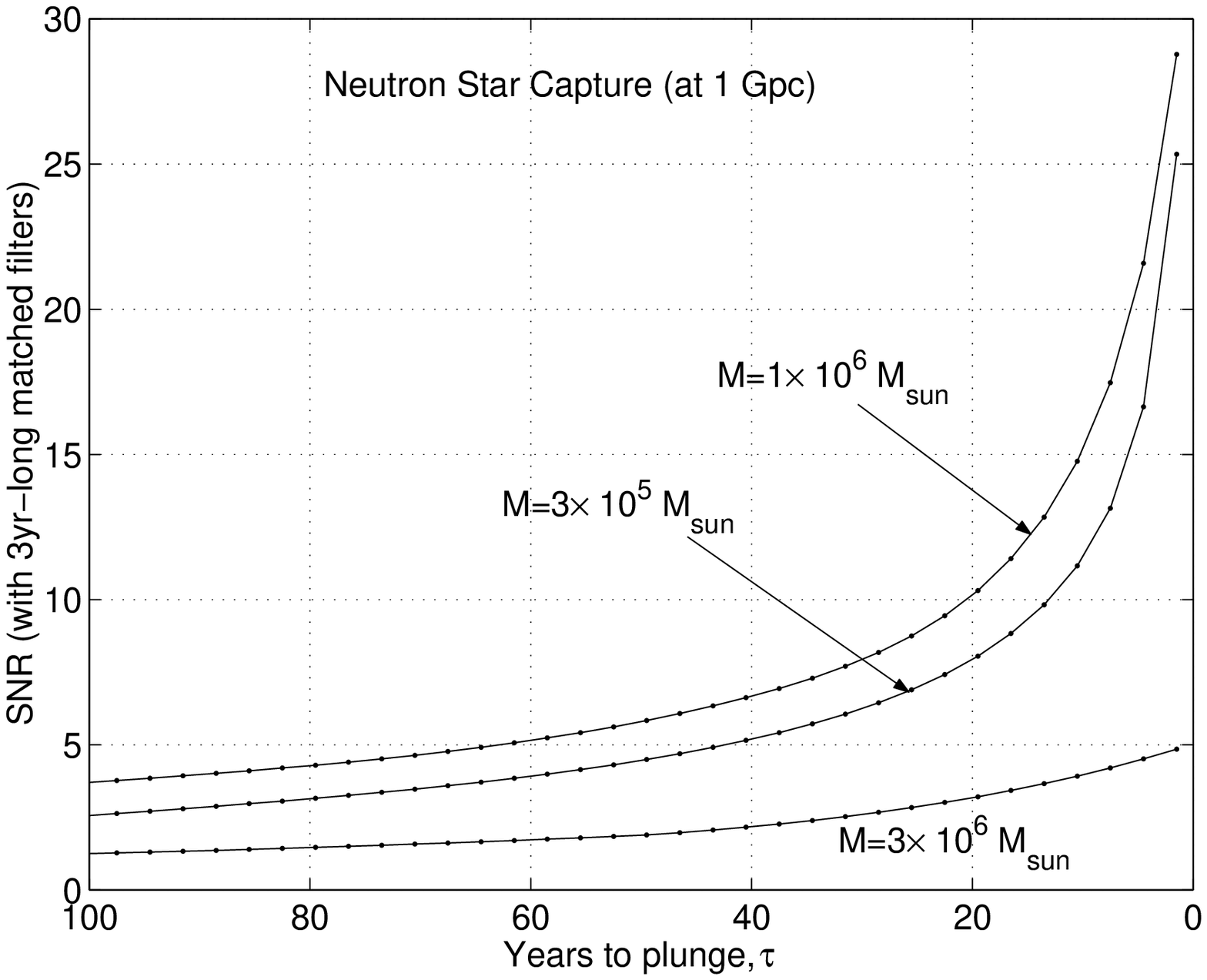}}
\centerline{\epsfysize 6cm \epsfbox{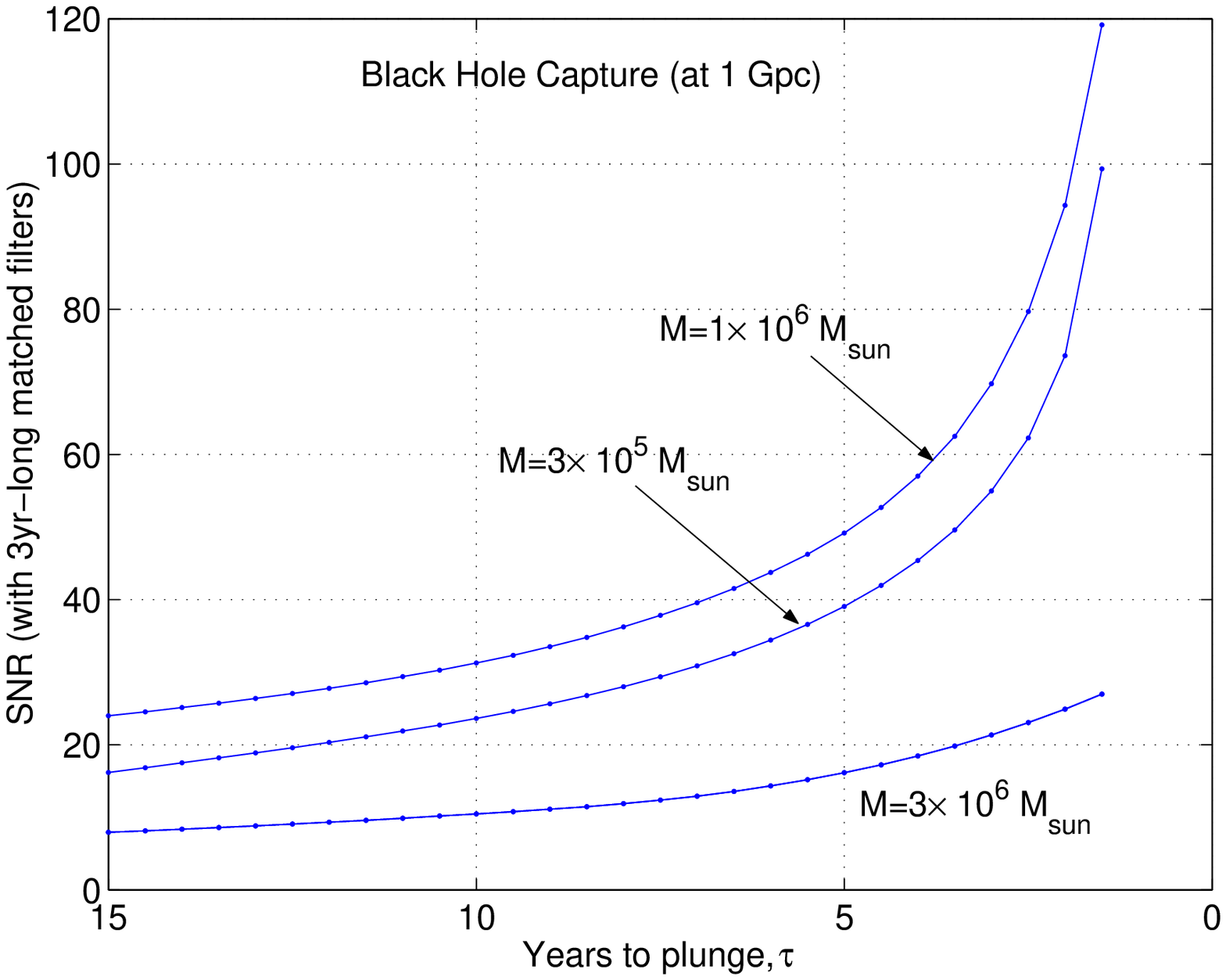}}
\caption{\protect\footnotesize
Combined SNR (from two synthetic Michelsons) as a function of time $\tau$ left to plunge for the three cases
shown in Fig.\ \ref{fig-Erad}, at a fiducial distance of 1 Gpc.
Each SNR data point represents the matched filter SNR for a 3-yr-long waveform segment
centered at $\tau$.
}
\label{fig-SNR}
\end{figure}

The second piece of input required for our estimate of subtractable noise
is the SNR output from typical sources, as a function of $\tau$.
Following Finn and Thorne~\cite{FT} we write the two-detector
(sky-averaged) SNR$^2$ as a sum of contributions from
all harmonics of the orbital frequency:
\be\label{sn2}
{\rm SNR}^2 =
2\, \Sigma_n\int\frac{h^2_{c,n}(f_n)}{f_n S_h (f_n)}\, d\ln f_n ,
\ee
\noindent where
\be\label{hcn}
h_{c,n} = (\pi d)^{-1}\sqrt{2\dot E_n/\dot f_n}
\ee
is the ``characteristic amplitude'',
$f_n = n\, \nu$, and $\dot E_n$ is the power radiated to infinity by GWs
at frequency $f_n$, given, to lowest order, by Eq.\ (\ref{dotEn}) above.
The results are shown in Fig.\ \ref{fig-SNR}.
In practice, we summed over the first $100$ harmonics. Also, we
performed the integral in the time domain, using
$\dot f_n = n\dot \nu$, with
$\dot \nu$ given by the lowest-order formula, Eq.~(\ref{pneqs1}).

Of course, the SNR results we estimate this way cannot be very accurate,
in general, since they are based on quasi-Newtonian orbits and
lowest-order radiation reaction formulae.
Their inaccuracy can be gauged, to some extent, by comparison with
exact (to numerical accuracy) values obtained by Finn and Thorne~\cite{FT}
for the case of circular, equatorial orbits.
For the last year of inspiral of a $10 {M_\odot}$ BH into a
$10^6 {M_\odot}$ Schwarzschild MBH, and
using essentially the same
LISA instrumental noise curve that we use, but neglecting confusion
noise, Finn and Thorne find the
(sky-averaged) value ${\rm SNR} = 72$~\cite{thorne_private} for one synthetic
Michelson. For the same case (and again neglecting confusion noise), our
method estimates SNR = 105 (again, for one synthetic Michelson).
Therefore our SNR is $45 \%$ too high for this case.
Unfortunately, for eccentric orbits, and even for circular orbits several
years before the final plunge, there is no ``gold standard''
that we can compare our SNR estimates to. We shall assume that
our SNR estimates are correct to within a factor $\sim 2$, and we shall
see that this accuracy is good enough for drawing our main conclusions.

In the analysis below we shall assume that ${\rm SNR}_{\rm thresh} = 30$,
so sources are detectable out to an average distance $d_{\rm det}$ of
\be\label{ddet}
d_{\rm det}/(1 {\rm Gpc}) = \text{(average 3-yr SNR @ 1 Gpc)}/30  ,
\ee
where the 3-yr SNR is calculated as described above.
However, given the various
approximations and uncertainties
(the rough nature of our SNR calculation, the uncertainty
in our estimate of ${\rm SNR}_{\rm thresh}$, the fact that typical
observation times might be $\sim 5$ or more years instead of $3$),
we shall also point out how our results below would change if
$d_{\rm det}$ were a factor two larger or smaller than our estimate.

\subsubsection{Subtracting the resolvable WD captures}

In Sec.\ \ref{SecIV} above we estimated the spectral noise density
${\cal S}_h^{\rm WDcapt}$ from all WD captures (see Fig.\ \ref{fig-SNR-WD}).
What portion of this noise is subtractable?

Consider first WD captures that are completed (i.e, the WD is
swallowed by the MBH) during the LISA mission lifetime ($\sim 3-5$ yr).
From Fig.~(\ref{fig-SNR}), we see that on average SNR$ \sim 9$ for such captures, so
from Eq.~(\ref{ddet}) we estimate $d_{\rm det} \sim 0.3$ Gpc.
Since the proper-motion distance\footnote{We use the proper-motion distance $d_M$ for
estimates here, as opposed to the luminosity distance or angular diameter distance, since
the number of sources closer than $d_M$ scales like $d_M^3$, assuming
that their number per co-moving volume is time-independent.}
at $z=1$ is 3.3 Gpc, and since the contribution from sources
at $z < 1$ is around $85\%$ of the contribution from all $z$ (see the discussion
in Sec.\ \ref{Cosmology} above), we estimate that a mere
$(0.3/3.3)\times 0.85\sim 8\% $ of the energy from the sources that plunge
during LISA's lifetime is subtractable.

Consider next the contribution to the confusion energy from captures
that are {\em not} completed during LISA's lifetime. Clearly, for these
captures (which have a smaller 3-yr SNR, and so must be
closer to us than $d_{\rm det} \sim 0.3$ Gpc to be detectable),
the portion of subtractable energy will be even
smaller than the above 8\%. Let us attempt to estimate
this portion, very roughly:
Examining Fig.\ \ref{fig-SNR}, we find that the 3-yr SNR from
typical WD captures drops by a factor $\sim 5$ over the last $\sim 100$
years of inspiral. Thus WD captures with $\sim 100$ years to live can
be detected out to an average $\sim 0.06$ Gpc. These detectable sources
contribute
$\sim 2\%$ of the capture noise (since this fraction scales as
the detection distance), so averaging $\sim 2\%$ and  $8\%$, we estimate
that of all the capture noise from sources with $0-100$ years until
final plunge,
$\sim 5\%$ is resolvable. According to Fig.\ \ref{fig-Erad}, however, these
sources contribute, on the average, only around $55 \%$ of the total confusion
energy---the rest is attributed to captures that have more than 100 years to
plunge when LISA observes.
From the sources with $> 100$ years to
go until plunge, we similarly estimate that $\sim 1\%$ of the capture noise
is resolvable.
Hence, we estimate the overall fraction of subtractable energy from
WD captures is roughly $0.55\times 5\% + 0.45 \times 1\% \approx 3\% $.

In conclusion, we estimate that $\sim 97\% $ of the capture noise
from WDs represents an irreducible confusion noise:
\begin{equation}\label{SconfWD}
S_h^{\rm WDcapt}(f) \approx 0.97\times {\cal S}_h^{\rm WDcapt}(f).
\end{equation}

If we were to assume that $d_{\rm det}$ is twice as large as estimated
above (say, because we have underestimated the signal strength, or
the low-frequency noise is significantly better than the design goal, or
the mission lifetime is much longer than 3 years, or because our
estimate of ${\rm SNR}_{\rm thresh}$ was too pessimistic), then the fraction of
subtractable energy would only increase to $6\% $.
Clearly, in either case it is a good estimate to simply take
$S_h^{\rm WDcapt}(f) \approx {\cal S}_h^{\rm WDcapt}(f)$.
Of course, this last approximation becomes even better
if $d_{\rm det}$ is smaller than estimated above.

\subsubsection{Subtracting the resolvable NS captures}

We next apply similar arguments to estimating the fraction
$S_h^{\rm NScapt}/{\cal S}_h^{\rm NScapt}$ of unsubtractable noise from
captures of NSs.
Fig.~13 shows that NS captures with $\tau$ less than a few years
can be detected to a distance $\sim 0.6$
Gpc--- roughly twice as far out as for WDs.
Therefore, the same reasoning as above for WDs suggests
that roughly $16\%$ of the
GW energy impinging on the detector from NS captures is resolvable
and hence subtractable
(i.e., twice the fraction we found for WDs).
Examining Figs.\ \ref{fig-SNR} and \ref{fig-Erad} again, we find that
captures that plunge $\sim 40$ years after LISA ends operations
are detectable to a distance of $\sim 0.17$ Gpc, so $\sim (.17/3.3)*0.85 \approx 4\%$
of the energy from these is subtractable.
Estimating that half the NS confusion background comes from sources with
$0-40$ years before plunge, and half from sources with $> 40$ years, and
estimating that the average subtractable noise portion for these
two classes are $10\%$ and $2\%$, respectively, we find (very roughly)
\begin{equation}\label{SconfNS}
S_h^{\rm NScapt}(f) \approx 0.94\times {\cal S}_h^{\rm NScapt}(f).
\end{equation}
If we were to assume NS captures are actually detectable to twice
the distance estimated above, we would still have
$S_h^{\rm NScapt}(f) \approx 0.88\times {\cal S}_h^{\rm NScapt}(f)$.
Clearly, as with the WDs, it is a good first approximation to simply take
$S_h^{\rm NScapt}(f) \approx {\cal S}_h^{\rm NScapt}(f)$.

\subsubsection{Subtracting the resolvable BH captures}

From Fig.~\ref{fig-SNR} we estimate that BH captures that plunge
during LISA's lifetime will have an average 3-yr SNR of $\sim 80$ at $1$ Gpc, and
so will be visible to $d_{\rm det} \sim 2.7$ Gpc (so almost to $z=1$).
Hence, unlike the situation with WDs and NSs,
most of the energy from
BH captures {\em that plunge during LISA's lifetime} is probably attributed
to resolvable sources, and hence is subtractable.

To estimate the confusion noise level we refer again
to Figs.\ \ref{fig-SNR} and \ref{fig-Erad}:
BH sources with $\sim 10$ years to go before plunge are detectable only out to
$\sim 0.7 $ Gpc, so $\sim 80\%$ of the background from these sources
is {\it not} subtractable. From Fig.\ \ref{fig-Erad},
roughly $40\%$ of the GW energy is released
more than 10 years before final plunge, and so
(averaging $80 \%$ and $100 \%$, and then multiplying by $0.4$) we
estimate that at least $\sim 35 \%$ of the BH background is unsubtractable:
\begin{equation}\label{SconfBH}
S_h^{\rm BHcapt}(f) \gtrsim 0.35\times {\cal S}_h^{\rm BHcapt}(f).
\end{equation}
(Note the RHS is a lower limit, since here we have not included
the hard-to-estimate contribution from unresolvable sources
with $\tau < 10$ yr.)
If we instead assume $d_{\rm det} = 1.4$ Gpc for BHs with $10$ years to
go before plunge (i.e, twice what we estimated above), then the
same steps yield
$S_h^{\rm BHcapt}(f) \gtrsim 0.3\times {\cal S}_h^{\rm BHcapt}(f)$.
On the other hand, if $d_{\rm det}$ is half what we estimated above,
then $\sim 40\% $ of the energy from BHs that plunge during LISA's
operation will not be subtractable, and for sources with
$\sim 10$ years to go before plunge, the unsubtractable fraction increases
to $90\% $. Thus overall, the unsubtractable portion would be
$\sim 75\% $ [$\approx 0.5(0.4+0.9)0.6  + 0.5(0.9+1.0)0.4$];
i.e., $S_h^{\rm BHcapt}(f) \approx 0.75\times {\cal S}_h^{\rm BHcapt}(f)$.

The above estimates were made assuming $S_h^{\rm BHcapt}(f)$ does not
significantly raise the total effective noise level.
However, if the BH capture
rates are at the high end of the estimated levels, then $S_h^{\rm
BHcapt}(f)$ {\it does} significantly raise the total noise curve, thereby
reducing the distance out to which the BH sources can be resolved.
Then $S_h^{\rm BHcapt}(f)$ would be in the intermediate regime shown
in Fig.~10, where it is difficult to estimate without going through
the entire recursive procedure described in Sec.\ \ref{Sec:scaling}.

For this reason, at the highest BH capture rates ($\kappa^{\rm BH}
\gtrsim 3 \times 10^{-7})$, our present work
is simply too crude to place very restrictive limits on the ratio
$S_h^{\rm BHcapt}(f)/{\cal S}_h^{\rm BHcapt}(f)$. Instead, we
quote the following range
\be\label{range}\label{SconfBH2}
S_h^{\rm BHcapt}(f) = (0.3 - 1 ) \times {\cal S}_h^{\rm BHcapt}(f)
\ee
(noting that the upper end of this range would  be approached
only if the BH capture rate is very high, so that capture noise
significantly raises the overall noise level)
and leave it to future work to improve this estimate.

\section{LISA's total noise curve} \label{SecVI}

\begin{figure}[htb]
\centerline{\epsfysize 10cm \epsfbox{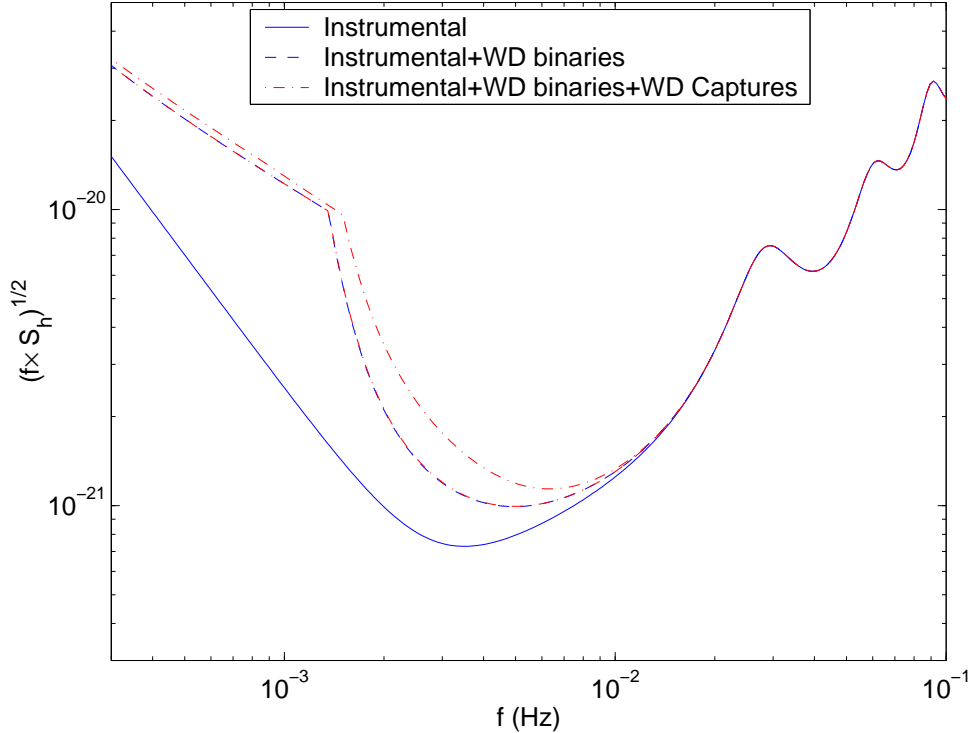}}
\caption{\protect\footnotesize
Total LISA noise (dash-dot line), including instrumental noise,
confusion noise from WD binaries, and confusion noise from captures of only
{\em White Dwarfs}.
The two total-noise curves correspond to the higher and lower end of
the estimated event rate for WD captures. (For the lower rate, the level of
WD capture confusion noise is low enough that the corresponding total-noise curve
appears to coincide in this plot with the dashed curve, representing
instrumental and WD-binaries confusion noise only.)
By virtue of Eq.\ (\ref{SconfWD}), we have assumed here that the {\em entire}
capture background ${\cal S}_h^{\rm WDcapt}$ contributes to the confusion noise
$S_h^{\rm WDcapt}$. The total noise curve has been calculated using Eq.\
(\ref{Stot}).
}
\label{fig-StotWD}
\end{figure}

\begin{figure}[htb]
\centerline{\epsfysize 10cm \epsfbox{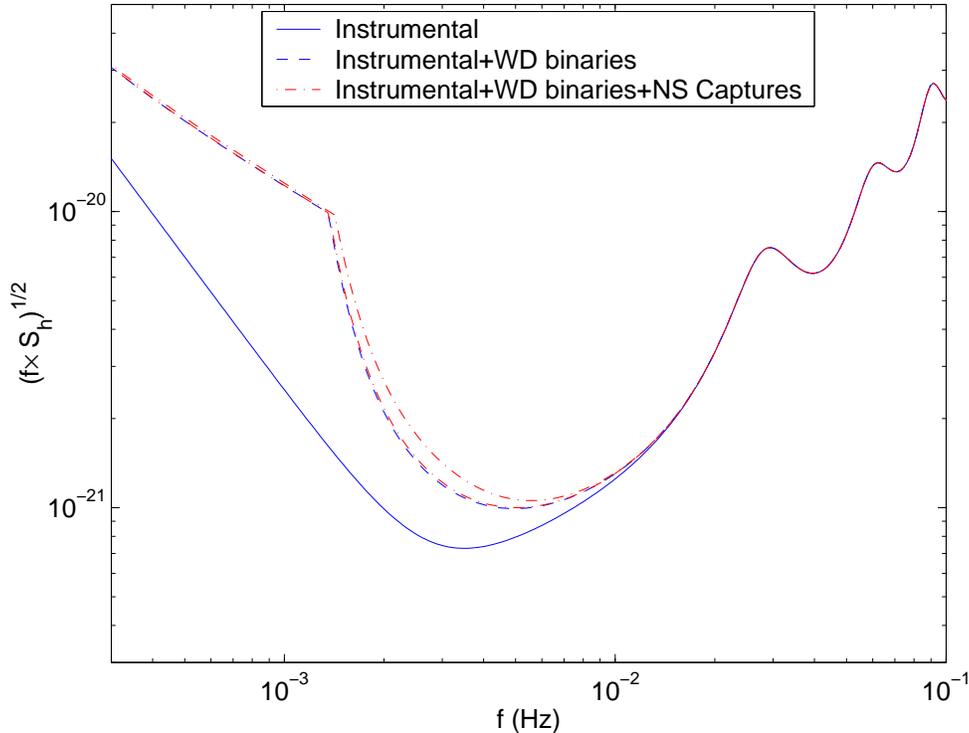}}
\caption{\protect\footnotesize
Same as in Fig.\ \ref{fig-StotWD}, but this time showing the total LISA noise
when capture confusion from only {\em Neutron Star} sources is taken into
account. By virtue of Eq.\ (\ref{SconfNS}), we have assumed here that the
{\em entire} background ${\cal S}_h^{\rm NScapt}$ contributes to the confusion
noise $S_h^{\rm NScapt}$.
}
\label{fig-StotNS}
\end{figure}

\begin{figure}[htb]
\centerline{\epsfysize 10cm \epsfbox{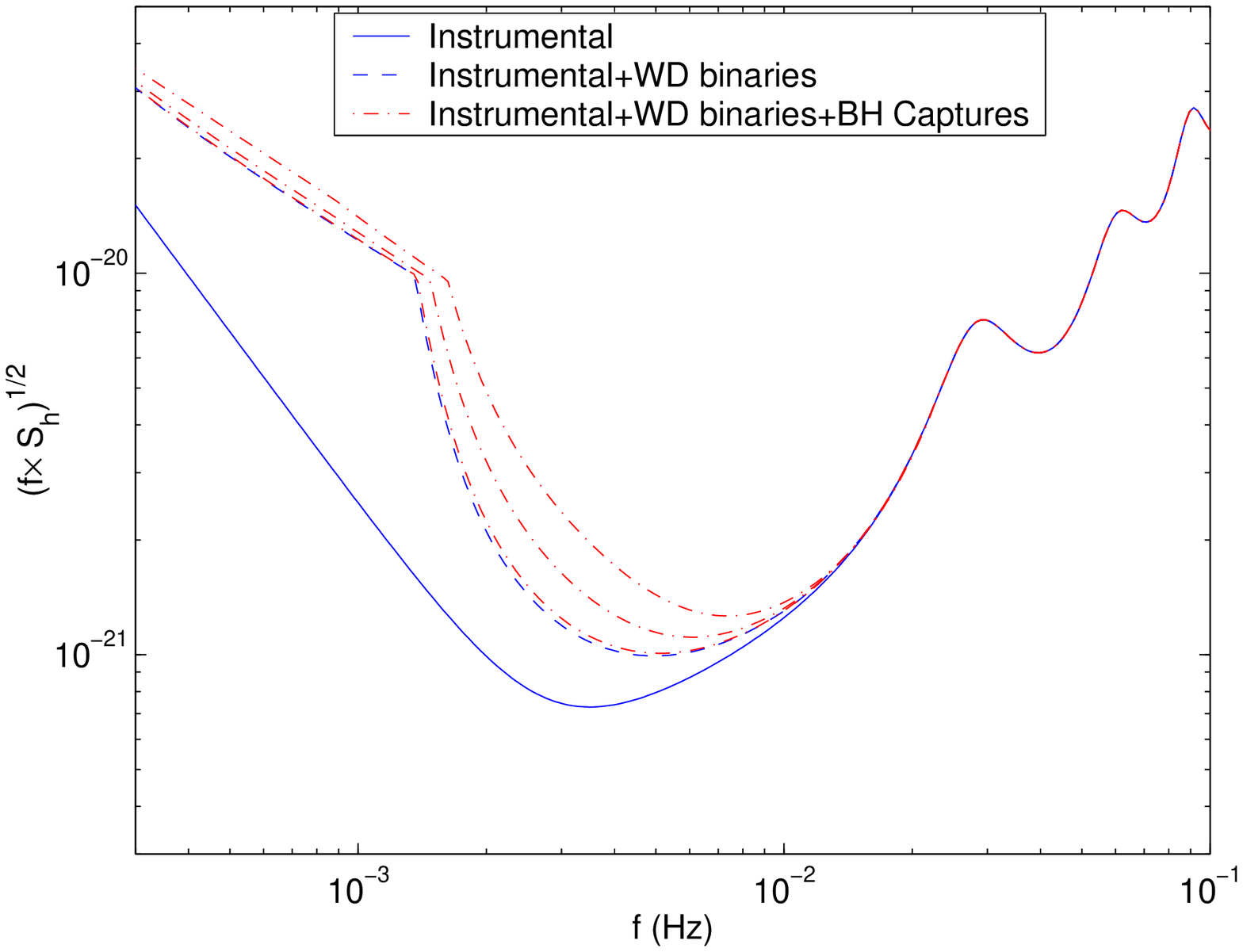}}
\caption{\protect\footnotesize
Same as in Fig.\ \ref{fig-StotWD}, but
with capture confusion from only {\em Black Hole} sources taken into account.
The three dash-dot lines show the total noise curve under different
assumptions as to the event rate $\kappa^{\rm BH}$ and the fraction of
the background that is subtractable.
The lower total-noise curve refers to the {\em lower}
end of event rate estimates ($\kappa^{BH} = 6 \times 10^{-8}$),
with $S_h^{\rm BHcapt}$ assumed to be $0.3\times {\cal S}_h^{\rm BHcapt}$.
The middle total-noise
curve corresponds to the {\em upper} end of rate estimates
($\kappa^{BH} = 6 \times 10^{-7}$), again with
$S_h^{\rm BHcapt} = 0.3\times {\cal S}_h^{\rm BHcapt}$.
The upper curve assumes the upper end of rate estimates, but with
$S_h^{\rm BHcapt}= {\cal S}_h^{\rm BHcapt}$.
Thus the upper curve represents an upper
limit on the effect of confusion noise from BH captures.
}
\label{fig-StotBH}
\end{figure}

Having estimated what fractions of the three CO backgrounds are
not subtractable, we may finally plot their effects on
LISA's total noise curve $S^{\rm eff}_h(f)$.
Figures \ref{fig-StotWD}--\ref{fig-StotBH} depict
$S^{\rm eff}_h(f)$, as derived from Eq.\ (\ref{Stot}), with the contributions
from the different CO species (WDs, NSs, or BHs) considered separately.
[Recall from the discussion around  Eq.~(\ref{Stot})
that in the
range $f \gtrsim 1$mHz, the effect of the capture confusion noise
(like the ones of instrumental noise and EGWDB background) is effectively
increased by a factor that counts the bandwidth lost when fitting the GWDBs.]
In the case of WDs and NSs we have included {\em all} capture noise as
confusion noise, since our above estimates suggest that the subtractable
portion of the noise would be small for these sources. For BH captures
we have assumed that $0.3$ of the capture noise is unsubtractable, but have
also shown the extreme case where $S_h^{\rm BHcapt}={\cal S}_h^{\rm BHcapt}$ [see
Eq.\ (\ref{SconfBH2})]. For each species we also refer to
two cases, corresponding to the lower and higher end of the estimated
event rate for that species. Note that the astrophysical event rate
remains the main source of uncertainty in our analysis, and clearly
overwhelms the uncertainty introduced by our crude estimate of the ratio
$S_h^{\rm capt}/{\cal S}_h^{\rm capt}$.


\section{Summary, implications, and future work}
\label{SecVII}

We have seen that astrophysical rates for CO captures are either near
(within one or two orders of magnitude) or at the
point where confusion noise from these sources begins to dominate
LISA's noise curve. That is, basically, a good situation: Such event
rates maximize the detection rate for these interesting sources.
Moreover, even at the high end of estimated rates, LISA's total noise
level $[f S^{\rm eff}_h(f)]^{1/2}$ is raised by less than a factor
$\sim 2$ by the CO capture confusion noise, so other LISA science
(such as detections of MBH-MBH mergers) is not jeopardized.

\begin{figure}[htb]
\centerline{\epsfysize 6cm \epsfbox{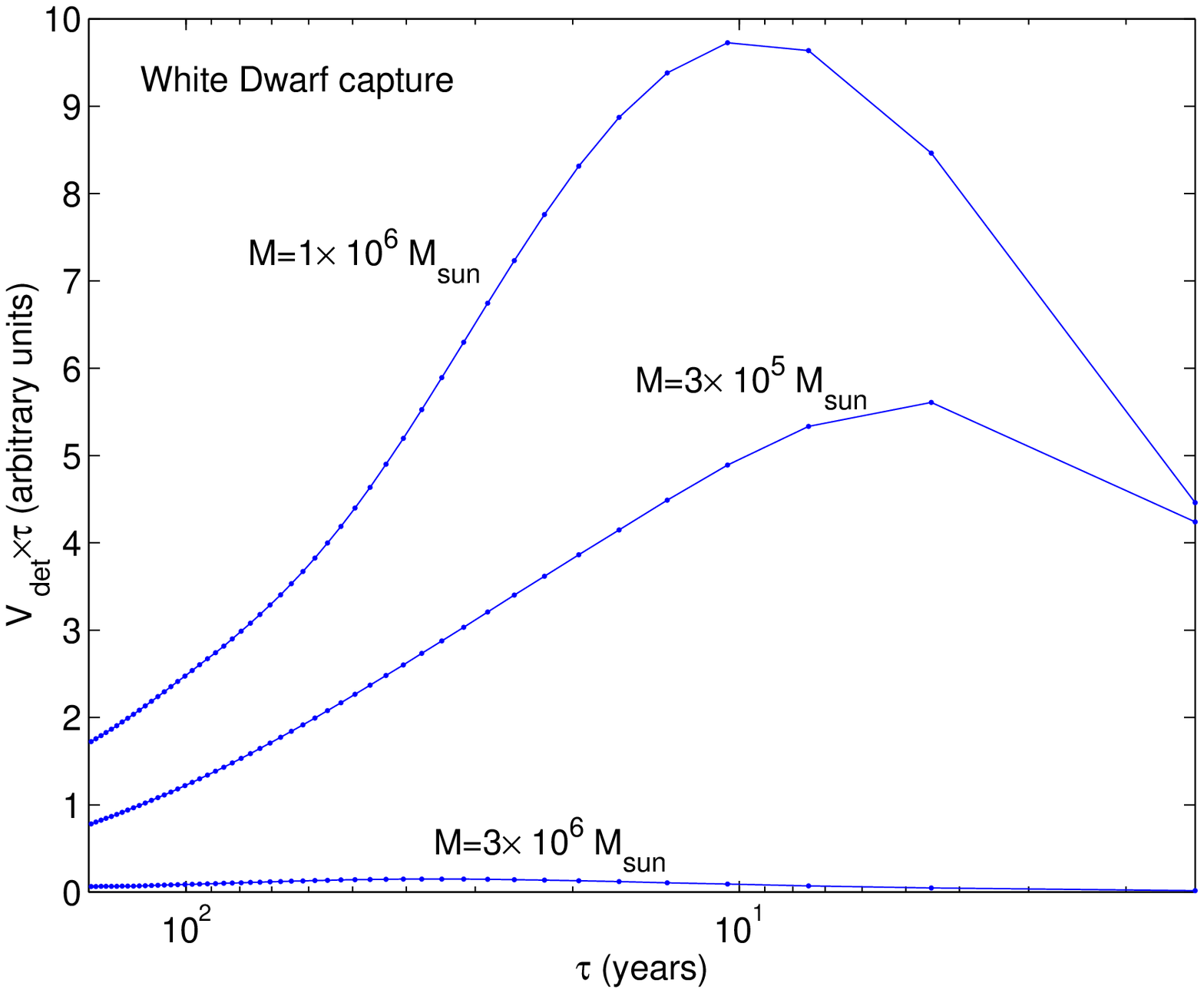}\hspace{4mm}
\epsfysize 6cm \epsfbox{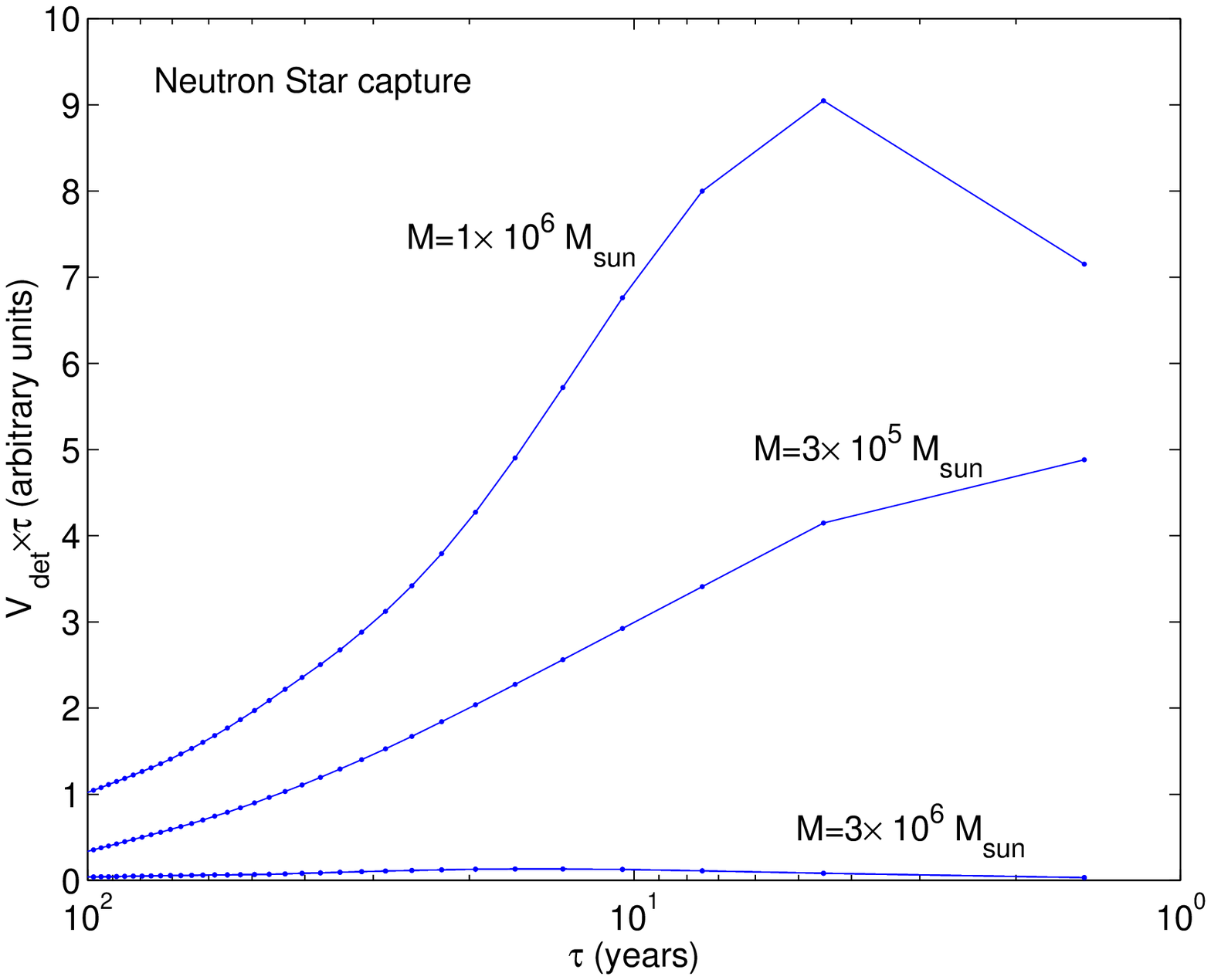}}
\centerline{\epsfysize 6cm \epsfbox{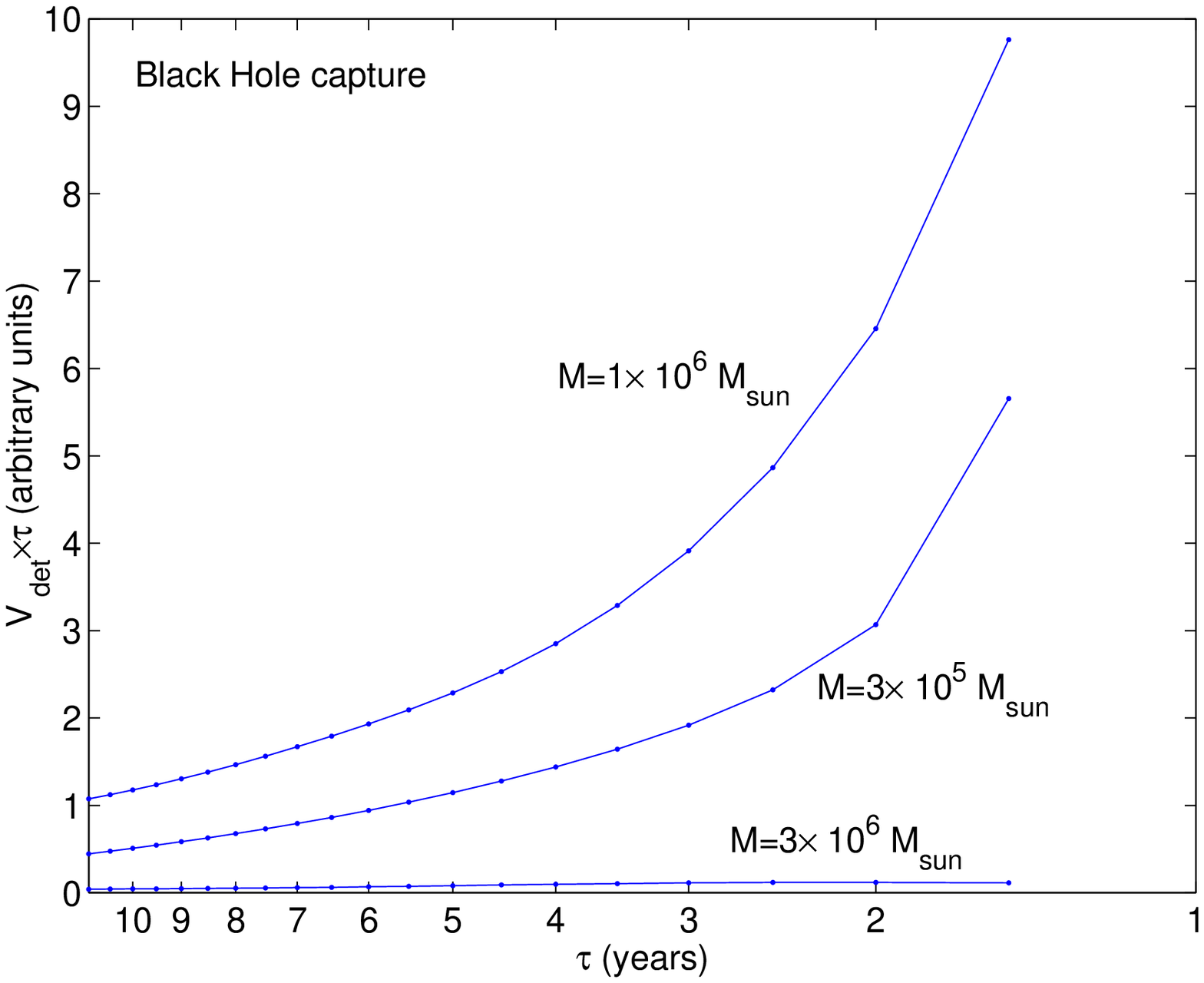}}
\caption{\protect\footnotesize
The detection volume $V_{\rm det}$ times $\tau$ (the time from the
middle of the 3-yr observation until plunge),
as a function  of $\ln\tau$, for our three fiducial sources.
The total number of detected captures is proportional to the
integral $\int{V_{\rm det} \tau \, d\ln\tau }$.
Roughly half (one-third) of the WDs (NSs) that LISA detects will have
$\tau > 10$yrs. BH detections will be dominated by sources with
$\tau < 3$ yr, but roughly $20\%$ will have $\tau > 5$yr.
}
\label{fig-VT}
\end{figure}

Let us also highlight an important point that is implicit in Fig.~13:
For WD capturess, HPOs will account for roughly half the detections.
(Again, HPOs are ``holding-pattern objects,'' i.e.,
COs detected $10$ or more years before they plunge).
To see this, in Fig.~\ref{fig-VT}
we plot (in arbitrary units) the detection volume $V_{\rm det}$
times the time until plunge $\tau$, versus $\ln\tau$, for our fiducial sources.
This uses the same information as contained in Fig.~13, since we
have simply taken $V_{\rm det} \propto d_{\rm det}^3 \propto ({\rm SNR} @ 1
{\rm Gpc})^3$; however, since
the total number of detected captures (for each type of source)
is proportional to $\int{V_{\rm det} \tau \, d\ln\tau}\,$, this
representation allows one tell at a glance the relative importance
of HPOs to the total detection rate.
In particular, we find that $\sim 1/2$ of detected WDs and $\sim 1/3$ of
detected NSs will have $\tau > 10 $ yr. While the BH detection rate will be
dominated by sources with $\tau\lesssim 3$ yr, $\sim 5-10\%$ of BH detections
will have $\tau > 10$ yr as well.
[This last estimate is based on a linear extrapolation of the curves in
Fig.~(\ref{fig-VT}) to larger $\tau$ values.
Unfortunately, at  $\tau \gtrsim 12$ yr, for the lower MBH mass, the BH
orbits attain very high eccentricities ($e > 0.9$), rendering our
code unreliable.]
Clearly, the results here are quite crude (most importantly, the
results in Figs.~13 and 14 are all based
on a single, fiducial value of the final eccentricity, $e_{\rm LSO} = 0.15$),
but still the moral seems to be clear: When searching over
the large parameter space of possible capture signals, it will be
worthwhile to construct template families that include
waveforms from HPOs. We also point out that many of the HPOs that LISA
detects will still be ``alive'' when some next-generation LISA is flown
(and, for some of the WDs, next-next-generation LISA, etc.) which may
allow for especially sensitive tests of GR in the future.

Clearly, the analysis in this paper has been crude in many ways. In
particular, most of the estimates in Sec.\ \ref{SecV} were based on
taking results from a few fiducial cases, and then ``averaging by
eye.''  Nevertheless, the uncertainties due to our approximations are
clearly dwarfed by the uncertainties in the astrophysical rates.
Moreover, our basic conclusion regarding WDs and NSs---that most of
the GW energy we receive from these captures is not resolvable by LISA
and so represents a confusion noise---seems very robust.
For the BH case, our estimated range of $S_h^{\rm capt}$ is rather
broader: $S_h^{\rm capt} \approx (0.3-1)\times {\cal S}_h^{\rm capt}$.
However, here too, the basic moral remains clear: To raise LISA's overall
effective noise level by even a factor $\sim 2$, the BH capture rate would
have to be at the high end of its estimated range, resulting in several
hundred detections per year---a compensation devoutly to be wished.

\acknowledgements
We thank the members of LIST's Working Group 1, and especially Sterl Phinney,
from whom we first learned of the problem considered in this paper.
We thank Marc Freitag for helpful discussions of his capture simulations,
and Markus P\"ossel for helping invent the term ``holding pattern object.''
We thank S. Hughes for pointing us to Ref [9].
C.C.'s work was partly supported by NASA Grant NAG5-12834.
L.B.'s work was supported by NSF Grant NSF-PHY-0140326 (`Kudu'),
and by a grant from NASA-URC-Brownsville (`Center for Gravitational
Wave Astronomy'). L.B. thanks the Albert Einstein
Institute, where part of this work was carried out, for its
hospitality.

\end{document}